\documentclass[aps,prb,twocolumn,showpacs,superscriptaddress]{revtex4-1}
\usepackage{graphicx,amssymb,amsfonts,amsmath,bbm}

\begin{document}
\newcommand{\AR}[1]{{\bf AR} {\it #1} {\bf END}}
\newcommand{\OLD}[1]{{\bf OLD} {\tiny #1}}
\newcommand{\sgn}{{\rm sign}}

\title{
Thermalization and dissipation in out of equilibrium quantum systems: A perturbative renormalization group approach}
\author{Aditi Mitra}
\affiliation{Department of Physics, New York University, 4 Washington
  Place, New York, New York 10003, USA}
\author{Thierry Giamarchi}
\affiliation{DPMC-MaNEP, University of Geneva, 24 Quai Ernest-Ansermet, CH-1211 Geneva, Switzerland}
\date{\today}


\begin{abstract}
A perturbative renormalization group approach is employed to study the effect of a
periodic potential on a system of one-dimensional bosons in
a non-equilibrium steady-state due to an initial interaction quench. The renormalization group
flows are modified significantly from the well known equilibrium Berezinski-Kosterlitz-Thouless form.
They show several new features such as, a generation of an effective temperature,
generation of dissipation, as well as a change in the location of the quantum critical point
separating the weak coupling and strong
coupling phases. Detailed results on the weak-coupling side of the phase diagram are presented,
such as the renormalization of the parameters and the asymptotic behavior of the correlation functions.
The physical origin of the generated temperature
and friction is discussed.
\end{abstract}


\maketitle

\section{Introduction}

The high degree of tunability and control associated with cold-atomic gases~\cite{Bloch08}
has motivated an explosion of theoretical activity involving the study of
dynamics of interacting quantum systems. Several interesting problems could be studied in
these systems such as quantum quenches (see Ref.~\onlinecite{Polkovrev09}
and references therein) and other classes of steady-state nonequilibrium phenomena such as
systems subjected to a time dependent noise.~\cite{Torre10} In all these situations one of
the fundamental
questions is what is the steady-state of the system, and in particular whether the
system can thermalize.

Not surprisingly in such a difficult problem, exactly solvable models and integrable systems
have proven to be a good playground to address
these issues. The study of quench dynamics in many integrable systems reveal that the large number of
conserved quantities in the system prevents the system from thermalizing. Instead the long time
behavior of the
system is characterized by time-dependent or time-independent nonequilibrium
states.~\cite{Barankov06,Yuzbashyan06}
Often after a time-averaging, the resultant steady-state can be
described by a generalized Gibbs ensemble (GGE) constructed from identifying the conserved
quantities of the
system.~\cite{Rigol07,Cazalilla06,Calabrese06,Barthel08,Kollar08,Meden10,Fioretto10}
However the generality
and applicability of the GGE remains under debate since
not all observables can be
described by it.~\cite{Kollath07,Gangardt08,Iucci09,Rossini10,Lancaster10,Silva11}

Besides a lack of complete understanding of the long time behavior of
integrable and even some exactly solvable models after a quench, the more complex question of
how the presence of non-trivial interactions that cause scattering
and/or break integrability affect the behavior is largely open.
Numerical studies on finite systems show that for large enough breaking of integrability,
thermalization
is associated with a change of the level statistics from Poisson to
Wigner-Dyson,~\cite{Rigol10} with the onset of thermalization occurring via
a two-step process, where in the first step the system is trapped for a long time in a
nonequilibrium
prethermalized state.~\cite{Berges04,Kehrein08,Rigol09a,Rigol09b,Kollar11}  At the same time, fluctuations in
a finite-size system also play an important role~\cite{Roux09,Biroli10} so that generalization of
numerical results to
systems in the thermodynamic limit may not be straightforward.

The aim of the current paper is to address some of these questions in precisely
this limit of long times and infinite system size where numerical studies are hard.
In particular we study an exactly solvable model which is in a nonequilibrium steady-state due to an
initial quench, and explore the stability of the resulting athermal state
to non-trivial interactions that generate mode-coupling.
The effects of mode-coupling will be treated within a perturbative renormalization group (RG) approach.
While there are many candidate models to study this physics, due to experimental relevance, and its
relative simplicity, we choose to study a one dimensional system of interacting bosons,
leading to the so called Luttinger liquid physics.~\cite{Giamarchibook}
The excitations of such a system can be represented by
density modes, which are essentially independent. On such a system,
quenches corresponding to a change of the interaction reveal a steady-state which
still has independent modes, but these are now characterized by a nonequilibrium distribution that does
not relax to a thermal state.~\cite{Cazalilla06,Iucci09,Perfetto06,Perfetto11,Dora11,Mitra11b}
Note that we use the term ``steady-state'' to reflect the fact that averages of various
physical observables reach
a time-independent value at long times after the quench.

We study the effect of a mode coupling term, such as one generated by a periodic potential,
on this steady-state. We assume that the periodic potential has been
switched on very slowly, so that in the absence of the initial quench, the system reaches the ground state
in the presence of
the periodic potential. The results are presented in the parameter regime where the periodic potential is (dangerously)
``irrelevant'' so that a perturbative RG approach remains valid at arbitrary length-scales.
We find that infinitesimally weak potentials or mode-coupling can generate
an effective-temperature and cause the system to \emph{asymptotically} thermalize.
In addition, and somewhat unexpectedly a dissipation is also generated. Thus we find that the effective
low energy theory at
long times after the quench is a quadratic theory of thermal bosons with a finite lifetime.
This asymptotic thermalization and dissipation
occurs because the low frequency and momentum modes can transfer energy to the higher energy modes
(which are gradually eliminated in the RG procedure). These high energy modes thus act as a bath,
providing thermalization and dissipation. The dissipation thus shares features with Landau
damping where plasmons acquire a finite lifetime.~\cite{Lancaster11}
It also shares similarities with turbulent systems that are known to exhibit the well known Kolmogorov
cascades where energy is passed down
from large scale structures to small scale structures.~\cite{Frisch}
We believe that this type of behavior and mechanism for thermalization,
unraveled in a controlled way on the particular system we study, is quite general.

A short account of some of these results was given in Ref.~\onlinecite{Mitra11b}.
In the present paper we give a detailed
derivation of the above results and in particular discuss the general derivation of the RG
equations using the Keldysh approach which may be adapted to study other types of out of
equilibrium bosonic systems. We build on the results of Ref.~\onlinecite{Mitra11b} to derive new results.
In particular we show that the RG flow can be significantly modified by different quench protocols.
We also examine the behavior close to the critical point where the
external potential becomes relevant. In this limit
various peculiarities such as singular behavior of the expansion in
spatial and temporal gradients of the bosonic field are encountered.
In addition we present a detailed discussion of the violation of the fluctuation dissipation
theorem of the post-quench system by studying the frequency dependence of two point correlation functions
showing that even though the system appears ``thermal'' in low energy scales, the crossover from low to
high energies is still complicated.

It should be noted that some RG based approaches to study quenches already exist in the literature.
For example the flow equation method was used to study quenches in a Fermi liquid~\cite{Kehrein08} and in the quantum sine-Gordon model,~\cite{Kehrein10} where for both cases a long-lived prethermalized regime was found. RG was also used to study the
a classical two-dimensional sine-Gordon model after a quench.~\cite{Mathey09} Here the effective temperature due to the quench was found to generate a dynamical vortex binding-unbinding transition, but unlike the quantum 1D problem studied in Ref.~\onlinecite{Mitra11b}
and this paper, no dissipation was generated. Furthermore, the system that we study also shares many
common features with a Luttinger liquid subjected to a nonequilibrium noise source,~\cite{Torre10}
where the particular form of the noise in our system arises due to the out of equilibrium occupation of the bosonic modes after the quench.
It is thus interesting to make a connection between the quench problem where in the asymptotic the system is free,
and systems where an external noise source is constantly imposed on the system,
and is found to exhibit similar properties.\cite{Torrelong}

The paper is organized as follows. In Section~\ref{model} the basic model and notation is introduced and
the equilibrium properties of the
relevant response and correlation functions are discussed. In Section~\ref{quench} the
interaction quench in the Luttinger liquid, and the properties of the resultant nonequilibrium steady-state is presented. In Section~\ref{rg} the periodic potential is introduced and the perturbative RG
equations are derived. In Section~\ref{sol} the solution of the RG equations are presented, and in
Section~\ref{action} the resultant low energy theory near the fixed point is discussed.
Finally in Section~\ref{summ} we present our conclusions and discuss open questions.

\section{Model} \label{model}

The Hamiltonian for interacting bosons in a periodic potential is
\begin{eqnarray}
&&H= H_0 + V_{sg}\\
&&H_0 = \frac{u}{2\pi}\int dx \left[K\left(\pi \Pi(x)\right)^2 + \frac{1}{K}\left(\partial_x \phi(x)\right)^2
\right]  \label{HSG}\\
&&V_{sg} = - \frac{gu}{\alpha^2}\int dx \cos\left(\gamma \phi(x)\right)
\end{eqnarray}
where $H_0$ is the quadratic part which describes the Luttinger liquid or
long lived sound modes. The density of these modes is $\rho  =-\partial_x\phi/\pi$, whereas
$\Pi= \partial_x \theta/\pi$ is the variable canonically conjugate to $\phi$.
$V_{sg}$ represents the periodic potential whose most important effect is a source of backscattering which can
localize the density modes via the well known Berezenskii-Kosterlitz-Thouless (BKT) transition.~\cite{Giamarchibook}

It is convenient to represent the  fields $\phi,\theta$ in terms of bosonic creation and annihilation operators
($b_p,b_p^{\dagger}$),~\cite{Giamarchibook}
\begin{eqnarray}
&&\phi(x) =  -(N_{R}+N_{L})\frac{\pi x}{L}
-\frac{i\pi}{L}\sum_{p\neq0}\left(\frac{L|p|}{2\pi}\right)^{1/2}\nonumber\\
&&\times \frac{1}{p}
e^{-\alpha|p|/2-ipx}\left(b_{p}^{\dagger} + b_{-p}\right), \\
&&\theta(x) = (N_{R}-N_{L})\frac{\pi x}{L} + \frac{i\pi}{L}\sum_{p\neq0}
\left(\frac{L|p|}{2\pi}\right)^{1/2}\nonumber \\
&&\times \frac{1}{|p|}e^{-\alpha|p|/2-ipx}\left(b_{p}^{\dagger} - b_{-p}\right).
\end{eqnarray}
where $\alpha^{-1}$ is an ultra-violet cutoff. Thus,
\begin{eqnarray}
H_0 = \sum_{p\neq 0} u|p|\eta_p^{\dagger} \eta_p
\end{eqnarray}
where
\begin{eqnarray}
\eta_{p} & = & \cosh\beta b_{p} + \sinh\beta b_{-p}^{\dagger},\\
\eta_{-p}^{\dagger} & = & \cosh\beta b_{-p}^{\dagger} + \sinh\beta b_{p}.
\end{eqnarray}
and $e^{-2\beta} = K, u = v_F/K$.

Since we are interested in nonequilibrium dynamics, we will use the
Keldysh formalism,~\cite{Kamenevrev} where
$\phi_{-/+}$ will denote fields that are (time/anti-time)-ordered on the Keldysh axis. Further, it will be convenient to
define quantum ($\phi_q$) and classical fields ($\phi_{cl}$),
\begin{eqnarray}
\phi_{\pm} = \frac{1}{\sqrt{2}}\left(\phi_{cl}\mp \phi_q\right)
\end{eqnarray}

\subsection{Correlation functions in equilibrium}

The two-point functions that are directly influenced by the periodic-potential, and therefore of interest to us are,
\begin{eqnarray}
&&C^{\pm,\pm}_{\phi\phi}(xt,yt^{\prime}) = \langle e^{i \gamma \phi_{\pm}(x,t)} e^{-i \gamma \phi_{\pm}(y,t^{\prime})}\rangle
\label{corrdef}
\end{eqnarray}
In order to compute the above correlators in the absence of a periodic potential ($g=0$), or
perturbatively in $g$, some useful identities are
\begin{eqnarray}
&&\langle \eta_{cl}(p,t)\eta_{cl}^{\dagger}(p^{\prime},t^{\prime})\rangle =
\delta_{pp^{\prime}} e^{-iu|p|(t-t^{\prime})}\coth\left(\frac{u|p|}{2T}\right)\\
&&\langle \eta_{cl}(p,t)\eta_{q}^{\dagger}(p^{\prime},t^{\prime})\rangle =
\delta_{pp^{\prime}} \theta(t-t^{\prime})e^{-iu|p|(t-t^{\prime})}\\
&&\langle \eta_{q}(p,t)\eta_{cl}^{\dagger}(p^{\prime},t^{\prime})\rangle =
-\delta_{pp^{\prime}} \theta(t^{\prime}-t)e^{-iu|p|(t-t^{\prime})}
\end{eqnarray}
where $T$ is the temperature of the bosons.

At $T=0$ and $g=0$ we find,
\begin{eqnarray}
&&C^{--}_{\phi\phi}(xt,yt^{\prime}) = \nonumber\\
&&e^{-\frac{\gamma^2K}{4}
\left[\ln\frac{\sqrt{ ((x-y)+u(t-t^{\prime}))^2 + \alpha^2}}{\alpha} +
\ln\frac{\sqrt{(x-y-u(t-t^{\prime}))^2 + \alpha^2}}{\alpha}\right]} \nonumber \\
&&e^{-\frac{\gamma^2K}{4}i {\rm sign}(t-t^{\prime})\left[
\tan^{-1}\frac{u(t-t^{\prime}) + x-y}{\alpha} + \tan^{-1}\frac{u(t-t^{\prime})-(x-y)}{\alpha}\right]}
\nonumber \\
\\
&&C^{++}_{\phi\phi}(xt,yt^{\prime}) = \nonumber \\
&&e^{-\frac{\gamma^2K}{4}
\left[\ln\frac{ \sqrt{(x-y+u(t-t^{\prime}))^2+ \alpha^2}}{\alpha} +
\ln\frac{\sqrt{(x-y-u(t-t^{\prime}))^2 +\alpha^2}}{\alpha}\right]} \nonumber \\
&&e^{\frac{\gamma^2K}{4}i{\rm sign}(t-t^{\prime})\left[
\tan^{-1}\frac{u(t-t^{\prime}) + x-y}{\alpha} + \tan^{-1}\frac{(u(t-t^{\prime})-(x-y))}{\alpha}\right]}
\nonumber \\
\\
&&C^{-+}_{\phi\phi}(xt,yt^{\prime}) \nonumber \\
&&= e^{-\frac{\gamma^2K}{4}
\left[\ln\frac{\sqrt{ (x-y+u(t-t^{\prime}))^2+ \alpha^2}}{\alpha} +
\ln\frac{\sqrt{ (x-y-u(t-t^{\prime}))^2 + \alpha^2}}{\alpha}\right]} \nonumber \\
&&e^{\frac{\gamma^2K}{4}i\left[\tan^{-1}\frac{ u(t-t^{\prime}) + x-y}{\alpha}
+ \tan^{-1} \frac{u(t-t^{\prime})-(x-y)}{\alpha}\right]}\nonumber \\
&&
\end{eqnarray}
Thus all correlators exhibit the typical power-law decay of a Luttinger-liquid, with an exponent $K_{eq}$ where
\begin{eqnarray}
K_{eq} = \frac{\gamma^2 K}{4} \label{Keqdef}
\end{eqnarray}

The oscillating factors in the above equations arise due to the Keldysh time-ordering and in
equilibrium have the right structure so that the well-known fluctuation-dissipation theorem (FDT) is obeyed.
To see this let us define the correlation function,
\begin{eqnarray}
&&C^K_{\phi\phi}(xt,yt^{\prime}) = \frac{-i}{2}\left(C^{--}_{\phi\phi}(xt,yt^{\prime})
+ C^{++}_{\phi\phi}(xt,yt^{\prime})\right)\label{CKdef}\\
&&=-ie^{-K_{eq}\left[\ln\frac{\sqrt{(x-y+u(t-t^{\prime}))^2+\alpha^2}}{\alpha} +
\ln\frac{\sqrt{(x-y-u(t-t^{\prime}))^2 + \alpha^2}}{\alpha}\right]} \nonumber \\
&&\cos\left[K_{eq}\tan^{-1}\left(\frac{u(t-t^{\prime}) + x-y}{\alpha} \right)\right. \nonumber \\
&&\left. + K_{eq}\tan^{-1}\left(\frac{u(t-t^{\prime})-(x-y)}{\alpha}\right)\right]
\end{eqnarray}
and the response function
\begin{eqnarray}
&&C^R_{\phi\phi}(xt,yt^{\prime}) = \frac{-i}{2}\left(C^{--}_{\phi\phi}(xt,yt^{\prime})
- C^{-+}_{\phi\phi}(xt,yt^{\prime})\right)\label{CRdef}\\
&&=-\theta(t-t^{\prime})\times \nonumber\\
&&e^{-K_{eq}\left[\ln\frac{\sqrt{(x-y+u(t-t^{\prime}))^2+\alpha^2}}{\alpha} +
\ln\frac{\sqrt{(x-y-u(t-t^{\prime}))^2 + \alpha^2}}{\alpha}\right]} \nonumber \\
&&\sin\left[K_{eq}\tan^{-1}\left(\frac{u(t-t^{\prime}) + x-y}{\alpha} \right)\right. \nonumber \\
&&\left. + K_{eq}\tan^{-1}\left(\frac{u(t-t^{\prime})-(x-y)}{\alpha}\right)\right]
\end{eqnarray}
In Fourier space $C(q,\omega) = \int_{-\infty}^{\infty} dt \int_{-\infty}^{\infty} dx e^{i\omega t - i q x}
C(x,t)$ are in general complicated to compute. However we will in the subsequent sub-section only highlight how the
FDT  works, so that it is easy to follow how it is violated in the post-quench situation.

\subsection{Temperature from the Fluctuation Dissipation Theorem}

The FDT implies that
\begin{eqnarray}
C^K_{\phi\phi}(q,\omega) &&= \coth\left(\frac{\omega}{2T}\right)2{\rm Im}\left[C^R_{\phi\phi}(q,\omega)\right]\\
&&\xrightarrow {T=0} {\rm sign}(\omega)2{\rm Im}\left[C^R_{\phi\phi}(q,\omega)\right]
\end{eqnarray}

Using $\ln(\alpha +i(ut\pm r)) = \ln(\sqrt{\alpha^2+(ut\pm r)^2}) + i\tan^{-1}\frac{ut\pm r}{\alpha}$ (placing the branch-cut of the
logarithm on the negative real axis), we may write
\begin{eqnarray}
&&C^K_{\phi\phi}(q,\omega) = -2i\int_0^{\infty}dt\int_{-\infty}^{\infty}dr\cos(\omega t)
\cos(qr)\times \nonumber \\
&&\frac{1}{2}
\left[e^{-K_{eq}\ln\frac{\alpha + i(ut+r)}{\alpha}
- K_{eq}\ln\frac{\alpha + i (ut-r)}{\alpha}} + c.c.\right]\\
&&2{\rm Im}[C^R_{\phi\phi}(q,\omega)] = 2i\int_0^{\infty}dt\int_{-\infty}^{\infty}dr\sin(\omega t)
\cos(qr)\times \nonumber \\
&&\frac{1}{2i}\left[e^{-K_{eq}\ln\frac{\alpha + i(ut+r)}{\alpha}
- K_{eq}\ln\frac{\alpha + i (ut-r)}{\alpha}} - c.c.\right]
\end{eqnarray}
Thus the FDT $C^K = 2{\rm sign}(\omega){\rm Im}[C^R]$ implies
\begin{eqnarray}
&&{\rm sign}(\omega) \int_0^{\infty}dt\int_{-\infty}^{\infty}dr\sin(\omega t)
\cos(qr)\times\nonumber \\
&&\frac{1}{2i}\left[e^{-K_{eq}\ln\frac{\alpha + i(ut+r)}{\alpha}
- K_{eq}\ln\frac{\alpha + i (ut-r)}{\alpha}} - c.c.\right]\nonumber \\
&&+ \int_0^{\infty}dt\int_{-\infty}^{\infty}dr\cos(\omega t)
\cos(qr)\times \nonumber \\
&&\frac{1}{2}\left[e^{-K_{eq}\ln\frac{\alpha + i(ut+r)}{\alpha}
- K_{eq}\ln\frac{\alpha + i (ut-r)}{\alpha}} + c.c.\right]\nonumber \\
&&=0
\end{eqnarray}
The above implies the following ought to be true,
\begin{eqnarray}
&&{\rm Re}\left[\int_0^{\infty}dt \int_{-\infty}^{\infty}dr \cos(qr) e^{-i|\omega| t}
\times \right. \nonumber \\
&&\left. e^{-K_{eq}\ln\frac{\alpha + i(ut+r)}{\alpha}
- K_{eq}\ln\frac{\alpha + i (ut-r)}{\alpha}}\right]=0
\end{eqnarray}
By analytically continuing $it \rightarrow \tau$, it is straightforward to see that the
expression in the square-brackets is purely imaginary thus proving the FDT at $T=0$.

In the next section, when we study the long time behavior of the response and correlation after
an interaction quench in the Luttinger liquid, we will find that the FDT is violated due to the appearance
of a new nonequilibrium exponent $K_{neq}$ which governs the power-law decay,
while the oscillating factors are still associated with the equilibrium exponent $K_{eq}$.

\section{Interaction quench in the Luttinger liquid: Properties of the quadratic theory} \label{quench}

Let us suppose that the system at time $t<0$ is a Luttinger liquid with
interaction parameter $K_0$ and velocity $u_0$, and therefore described by the Hamiltonian
\begin{eqnarray}
&&H_i = \frac{u_0}{2\pi}\int dx
\left[K_0\left(\pi \Pi(x)\right)^2 + \frac{1}{K_0}\left(\partial_x \phi(x)\right)^2
\right]  \\
&&= \sum_{p\neq 0} u_0 |p| \eta_p^{\dagger} \eta_p
\label{Hidef}
\end{eqnarray}
We will consider the case where at $t=0$ there is an interaction quench from $K_0\rightarrow K$
so that the time evolution from $t >0$ is due to
\begin{eqnarray}
&&H_f = \frac{u}{2\pi}\int dx
\left[K\left(\pi \Pi(x)\right)^2 + \frac{1}{K}\left(\partial_x \phi(x)\right)^2
\right] \nonumber \\
&&= \sum_{p\neq 0}u|p|\gamma_p^{\dagger}\gamma_p
\label{Hfdef}
\end{eqnarray}
To preserve Galilean invariance (which is not necessary for the formalism), we assume
$u = v_F/K, u_0=v_F/K_0$. Note that,
\begin{eqnarray}
\begin{pmatrix} b_p \\ b_{-p}^{\dagger}\end{pmatrix}
= \begin{pmatrix} \cosh\beta & -\sinh\beta \\-\sinh\beta &\cosh\beta
\end{pmatrix}
\begin{pmatrix} \gamma_p \\ \gamma_{-p}^{\dagger}\end{pmatrix}\\
\begin{pmatrix} b_p \\ b_{-p}^{\dagger}\end{pmatrix}
= \begin{pmatrix} \cosh\beta_0 & -\sinh\beta_0 \\-\sinh\beta_0 &\cosh\beta_0
\end{pmatrix}
\begin{pmatrix} \eta_p \\ \eta_{-p}^{\dagger}\end{pmatrix}
\end{eqnarray}
where $e^{-2\beta_0}=K_0, e^{-2\beta} = K$. The quench for $K_0=1$ was studied in
Ref.~\onlinecite{Iucci09}, and more general interaction
quenches were studied in Refs.~\onlinecite{Perfetto11}
and~\onlinecite{Mitra11b}. However,
the distinction between response and correlation functions
were only first identified in Ref.~\onlinecite{Mitra11b}.
Here we give more details of the results that appear in Ref.~\onlinecite{Mitra11b},
and in addition discuss the crossover behavior from low to high frequencies.

Let us define the functions
\begin{eqnarray}
f(pt) &&= \cos(u |p| t)\cosh\beta_0 \nonumber \\
&&-i \sin(u |p| t)\cosh(2\beta-\beta_0)\\
g(pt) &&= \cos(u |p| t)\sinh\beta_0 \nonumber \\
&&+ i \sin(u |p| t)\sinh(2\beta-\beta_0)
\end{eqnarray}
which determine the time-evolution after the quench ($t > 0$),
\begin{eqnarray}
b_p^{\dagger}(t) + b_{-p}(t) &&=\left(f^*(pt)-g(pt)\right)\eta_p^{\dagger}(0) \nonumber \\
&&+
\left(f(pt)-g^*(pt)\right)\eta_{-p}(0)\\
b_p^{\dagger}(t) - b_{-p}(t) &&=\left(f^*(pt)+g(pt)\right)\eta^{\dagger}_p(0) \nonumber \\
&&-\left(f(pt)+g^*(pt)\right)\eta_{-p}(0)
\end{eqnarray}
Using the above,
the basic expectation value for the $\phi$-fields after the quench can be easily worked out to give,
\begin{eqnarray}
&&-i\langle \phi(xt) \phi(y t^{\prime})\rangle \xrightarrow{t+t^{\prime}\rightarrow \infty}\nonumber \\
&&= -\frac{i}{4}\int_{-\infty}^{\infty}\frac{dp}{|p|}e^{-\alpha |p|}\cos(p(x-y))\times \nonumber \\
&&\left[\frac{K_0}{2}\left(1+\frac{K^2}{K_0^2}\right)\cos{u|p|(t-t^{\prime})}
\coth\frac{u|p|}{2T}\right. \nonumber \\
&&\left.-i K \sin{u|p|(t-t^{\prime})}\right]\label{phphbasic}
\end{eqnarray}
where $T$ denotes the temperature of the Luttinger-liquid before the quench.
Above, terms that oscillate as $e^{-iu |p|(t+t^{\prime})}$ have been dropped. This is because such oscillating terms
give an over-all decay to the correlators of interest defined in Eq.~(\ref{corrdef}). Since we are ultimately interested in the long-time
limit rather than the transients, these terms are not important for us. Further, this approximation also leads to a significant simplification as one is now dealing with a nonequilibrium steady-state problem, defined by the following basic
retarded and Keldysh Green's functions,
\begin{eqnarray}
G_R(xt,yt^{\prime})
&&= -i\theta(t-t^{\prime})\langle\left[\phi(xt),\phi(yt^{\prime})\right]\rangle\nonumber \\
&&=-i\langle \phi_{cl}(xt)
\phi_q(yt^{\prime})\rangle\\
G_A(xt,yt^{\prime})
&&= i\theta(t^{\prime}-t)\langle\left[\phi(xt),\phi(yt^{\prime})\right]\rangle\nonumber \\
&&=-i\langle \phi_{q}(xt)
\phi_{cl}(yt^{\prime})\rangle\\
G_K(xt,yt^{\prime}) &&= -i\langle \{\phi(xt),\phi(yt^{\prime})\}\rangle \nonumber \\
&&= -i \langle
\phi_{cl}(xt)
\phi_{cl}(yt^{\prime})\rangle
\end{eqnarray}
which in Fourier space acquire the following form at $T=0$
\begin{eqnarray}
&&G_R(q,\omega) = \frac{\pi K u}{\left(\omega + i\delta\right)^2 - u^2 q^2}\nonumber \\
&&= \frac{\pi K}{2|q|}\left[\frac{1}{\omega -u |q| + i\delta} - \frac{1}{\omega + u |q| + i\delta}\right]
\label{GR}\\
&&G_K(q,\omega) = -i\frac{\pi^2}{2} K_0\left(1 + \frac{K^2}{K_0^2}\right)
\frac{{\rm sign}(\omega)}{|q|} \times \nonumber \\
&&\left[\delta(\omega - u |q|) - \delta(\omega + u |q|)\right]\label{GK}\\
&&= \frac{K_0}{2K}\left(1 + \frac{K^2}{K_0^2}\right) {\rm sign}(\omega)\left[G_R - G_A\right]\label{fdtg}
\end{eqnarray}
Note that the retarded Green's function $G_R$ depends only on the final Hamiltonian, and therefore is
not sensitive to the quench. The Keldysh Green's function $G_K$ on the other hand contains information about the
occupation probabilities of the bosonic modes, which can be far from thermal equilibrium due to the quench.
Thus $G_K$ depends on
the properties of both the initial Hamiltonian (via $K_0$) and the final Hamiltonian (via $K$).
In equilibrium $K=K_0$, and the FDT (for $T=0$) $G_K = {\rm sign}(\omega)\left[G_R - G_A\right]$ is recovered.

\subsection{Correlation functions after the quench}
It is convenient to define a nonequilibrium exponent,
\begin{eqnarray}
K_{neq}=\frac{\gamma^2}{8}K_0(1+K^2/K_0^2) \label{Kneqdef}
\end{eqnarray}
which we will show below represents the new power-law decay of the correlations in Eq.~(\ref{corrdef}).
Interestingly both the equilibrium exponent $K_{eq}$
defined in Eq.~(\ref{Keqdef}) and the exponent $K_{neq}$ affect the unequal time correlators, which
are found to be,
\begin{eqnarray}
&&C^{--}_{\phi\phi}(xt,yt^{\prime}) = \nonumber\\
&&e^{-K_{neq}
\left[\ln\frac{\sqrt{ ((x-y)+u(t-t^{\prime}))^2 + \alpha^2}}{\alpha} +
\ln\frac{\sqrt{(x-y-u(t-t^{\prime}))^2 + \alpha^2}}{\alpha}\right]} \nonumber \\
&&e^{-iK_{eq} {\rm sign}(t-t^{\prime})\left[
\tan^{-1}\frac{u(t-t^{\prime}) + x-y}{\alpha} + \tan^{-1}\frac{u(t-t^{\prime})-(x-y)}{\alpha}\right]}
\nonumber \\
\\
&&C^{++}_{\phi\phi}(xt,yt^{\prime}) = \nonumber \\
&&e^{-K_{neq}
\left[\ln\frac{ \sqrt{(x-y+u(t-t^{\prime}))^2+ \alpha^2}}{\alpha} +
\ln\frac{\sqrt{(x-y-u(t-t^{\prime}))^2 +\alpha^2}}{\alpha}\right]} \nonumber \\
&&e^{iK_{eq}{\rm sign}(t-t^{\prime})\left[
\tan^{-1}\frac{u(t-t^{\prime}) + x-y}{\alpha} + \tan^{-1}\frac{(u(t-t^{\prime})-(x-y))}{\alpha}\right]}
\nonumber \\
\\
&&C^{-+}_{\phi\phi}(xt,yt^{\prime}) \nonumber \\
&&= e^{-K_{neq}
\left[\ln\frac{\sqrt{ (x-y+u(t-t^{\prime}))^2+ \alpha^2}}{\alpha} +
\ln\frac{\sqrt{ (x-y-u(t-t^{\prime}))^2 + \alpha^2}}{\alpha}\right]} \nonumber \\
&&e^{iK_{eq}\left[\tan^{-1}\frac{ u(t-t^{\prime}) + x-y}{\alpha}
+ \tan^{-1} \frac{u(t-t^{\prime})-(x-y)}{\alpha}\right]}\nonumber \\
&&
\end{eqnarray}
The above agrees with the equal time ($t=t^{\prime}$)
correlators for $K_0=1$ studied in Ref~\onlinecite{Iucci09}.

The power-law decay is determined by $K_{neq}$ which
is a memory dependent exponent as it explicitly depends on the interaction parameter $K_0$
before the quench. On the other hand, $K_{eq}$ depends only on the interaction parameter of
the final Hamiltonian,
and characterizes the equilibrium $T=0$ properties of the system.
Further, $K_{neq} > K_{eq}$, so that the power-law
decay is always somewhat faster in the nonequilibrium steady-state. The faster decay occurs both for
$\langle e^{i\phi(1)} e^{-i\phi(2)}\rangle$ and the dual $\langle e^{i\theta(1)} e^{-i\theta(2)}\rangle$.
In that sense,
the effect of a quench is similar to a temperature, however the system remains in a critical state.
In the next section, we will find that
as a consequence of this, the periodic potential is always less relevant for the
nonequilibrium steady-state problem,
with the critical point shifting to smaller values of $K$. Similar power-law decays with nonequilibrium
exponents can also arise in open systems subjected to a nonequilibrium noise
source such as 1/f noise.~\cite{Torre10}

It is interesting to observe that two different quench protocols can lead to the same nonequilibrium steady-state,
at least
for a case where the steady-state is determined by the behavior of the $C_{\phi\phi}$ correlators.
To see this, for simplicity set $\gamma=2$ (so that $K_{eq}=K$). Then,
\begin{eqnarray}
K_{neq} = K  + \frac{(K-K_0)^2}{2K_0} \label{double}
\end{eqnarray}
From the above equation one may see that two different $K_0$ may lead to the same $K_{eq}$ and $K_{neq}$.
For example
$K_{eq}=1,K_{neq}=2$ can be obtained for a quench from
$K_0\rightarrow K$ where $K=1$, whereas the initial interaction
parameter $K_0$ can take two different values $K_0 = 2 \pm \sqrt{3}$. This behavior simply reflects the
fact that when
$K=K_0$, the system being in equilibrium, $K_{neq}=K$. On the other hand $K_{neq}$ increases with
respect to $K$ for
both types of quenches, one where $K_0 > K$ and the other where $K_0 < K$. For both these cases, the system
is driven out of
equilibrium, giving rise to a faster decay than in equilibrium.

As before, we now discuss the FDT ratio for the $C_{\phi\phi}$ two-point functions.
(Note that often our convention will be
to express length-scales in units of $\alpha$ and energy-scales in units of
$u/\alpha$).
The Keldysh correlation function (defined in Eq.~(\ref{CKdef})) is found to be
\begin{eqnarray}
&&C^K_{\phi\phi}(r,t) = -i\cos\left[K_{eq}\tan^{-1}\left(\frac{ut+r}{\alpha}\right)\right. \nonumber \\
&&\left. + K_{eq}\tan^{-1}\left(\frac{ut-r}{\alpha}\right)
\right]\times \nonumber \\
&&\left(\sqrt{\frac{\alpha^2}{\alpha^2+(ut-r)^2}}\right)^{K_{neq}}
\left(\sqrt{\frac{\alpha^2}{\alpha^2+(ut+r)^2}}\right)^{K_{neq}}
\end{eqnarray}
and the retarded correlation function (defined in Eq.~(\ref{CRdef})) is found to be,
\begin{eqnarray}
&&C^R_{\phi\phi}(r,t) = -\theta(t)\sin\left[K_{eq}\tan^{-1}\left(\frac{ut+r}{\alpha}\right)
\right. \nonumber \\
&&\left.+ K_{eq}\tan^{-1}\left(\frac{ut-r}{\alpha}\right)
\right]\nonumber \\
&&\left(\sqrt{\frac{\alpha^2}{\alpha^2+(ut-r)^2}}\right)^{K_{neq}}
\left(\sqrt{\frac{\alpha^2}{\alpha^2+(ut+r)^2}}\right)^{K_{neq}}
\end{eqnarray}

In general the Fourier transform of the above expressions for $C^{K,R}$ need to
be calculated numerically, and were briefly discussed in Ref.~\onlinecite{Mitra11b}.
In the $q=0,\omega \rightarrow 0$
limit however, analytic expressions for $C_{\phi\phi}$ can again be obtained. In particular
\begin{eqnarray}
C^K_{\phi\phi}(q=0,\omega=0) = -i I_{T0}
\end{eqnarray}
where (setting $u=1$)
\begin{eqnarray}
&&I_{T0} =\nonumber \\
&&2\alpha^2 \left[\frac{\pi}{2^{K_{neq}-1}(K_{neq}-1)
B\left(\frac{K_{eq}+K_{neq}}{2},\frac{K_{neq}-K_{eq}}{2}\right)}\right]^2\nonumber \\
&&\label{It0}
\end{eqnarray}
$B(x,y)$ being the beta function.
Similarly, we find
\begin{eqnarray}
2{\rm Im}\left[C^R_{\phi\phi}(q=0, \omega \rightarrow 0)\right]= -i\omega I_{\eta 0}
\end{eqnarray}
where (setting $u=1$)
\begin{eqnarray}
&&I_{\eta 0} = \nonumber \\
&&\alpha^3\left[\frac{\pi}{2^{K_{neq}-1}(K_{neq}-1)
B\left(\frac{K_{eq}+K_{neq}}{2},\frac{K_{neq}-K_{eq}}{2}\right)}\right]\nonumber \\
&&\times \left[\frac{\pi}{2^{K_{neq}-2}(K_{neq}-2)
B\left(\frac{K_{eq}+K_{neq}-2}{2},\frac{K_{neq}-K_{eq}}{2}\right)}  \right. \nonumber \\
&&\left. - \frac{\pi}{2^{K_{neq}-2}(K_{neq}-2)
B\left(\frac{K_{eq}+K_{neq}}{2},\frac{K_{neq}-K_{eq}-2}{2}\right)}\right]\nonumber \\
\label{Ieta0}
\end{eqnarray}
Note that in equilibrium $K_{eq}=K_{neq}$, and $I_{T,\eta 0}=0$.

\subsection{Violation of the quantum FDT and zero-frequency effective-temperature}

Even though the quantum FDT is not obeyed, one may define an effective-temperature $T_{eff,0}$  in the
low-frequency limit as follows,
\begin{eqnarray}
\frac{C^K_{\phi\phi}(q=0,\omega=0)}{2{\rm Im}\left[C^R_{\phi\phi}(q=0, \omega \rightarrow 0)\right]}
= \frac{2T_{eff,0}}{\omega}
\end{eqnarray}
where we find that the effective-temperature (in dimensions of $u/\alpha$) is
\begin{eqnarray}
{T}_{eff,0}= \frac{\alpha I_{T0}}{2 I_{\eta 0}}=\frac{K_{neq}-2}{2 K_{eq}}\label{teff0def}
\end{eqnarray}
As is typical of nonequilibrium systems, this effective-temperature depends on the correlation
function being studied as it certainly does not characterize the low
frequency properties of the simpler correlators in Eq.~(\ref{fdtg}).
Moreover, this temperature or equivalently the noise correlator
has a complicated frequency dependence.
However, as we will show by doing RG, the low-frequency limit of
the noise or the temperature has important physical consequences, as it changes the
long-time and distance behavior of
correlation functions by causing them to decay exponentially fast (rather than as a power-law with
exponent $K_{neq}$).

\begin{figure}
\includegraphics[totalheight=5cm]{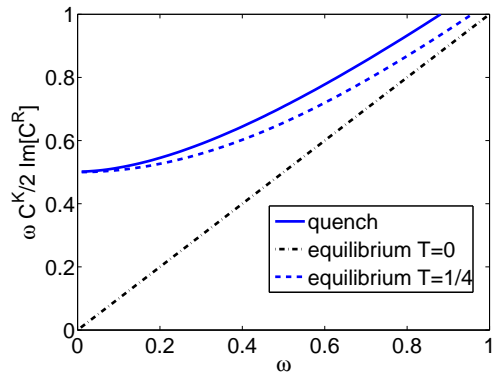}
\caption{\label{fdt1}The ratio $\frac{\omega C^K(q=0,\omega)}{2{\rm Im}[C^R(q=0,\omega)]}$
for a quench where $K_{eq}=2,K_{neq}=3$. This is compared with the equilibrium expression $\omega\coth\frac{\omega}{2T}$.}
\end{figure}

The crossover from $\omega \ll T_{eff,0} $ to $\omega \gg T_{eff,0}$ is illustrated in Fig.~\ref{fdt1}
which plots the ratio
$\frac{\omega C^K_{\phi\phi}(q=0,\omega)}{2{\rm Im}C^R_{\phi\phi}(q=0,\omega)}$. In equilibrium,
this ratio takes the value of
$\omega\coth\frac{\omega}{2T_{eff,0}}$ which in the high-frequency limit
becomes $\omega\coth\frac{\omega}{2T_{eff,0}}\xrightarrow{T_{eff,0} \ll \omega}|\omega|$, and
in the low-frequency
limit is $\omega\coth\frac{\omega}{2T_{eff,0}}\xrightarrow{T_{eff,0} \gg \omega}2T_{eff,0}$.
The plot shows that the nonequilibrium system shows a slower crossover to $|\omega|$ with increasing frequencies
than the equilibrium system, indicating that the occupation of higher energy modes decays slower
than exponential.
It should be noted that similar noise with a complicated cross-over behavior from low to high frequencies
was studied in open and driven systems near quantum critical points.~\cite{Mitra06,Mitra08a,Mitra08b,Takei10}
The low-frequency limit of the noise was found to cut off the power-law decay of critical fluctuations,
and to also cause a classical ordering-disordering phase transition.~\cite{Mitra06,Mitra08a}

\begin{figure}
\includegraphics[totalheight=5cm]{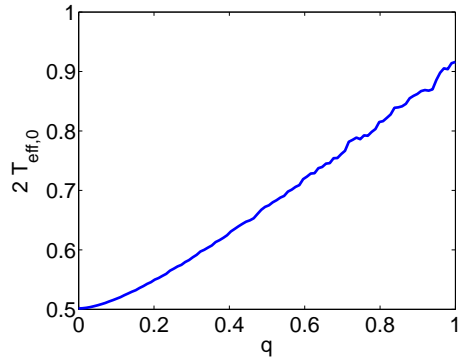}
\caption{\label{fdt2}$q$ dependence of $2T_{eff,0}$ for a quench where $K_{eq}=2$ and $K_{neq}=3$.}
\end{figure}

Another measure of the violation of FDT is to extract the momentum dependence of the zero-frequency
temperature, $2T_{eff,0}(q) = \frac{\omega C^K_{\phi\phi}(q,\omega=0)}{2{\rm Im}C^R_{\phi\phi}(q,\omega\rightarrow 0)}$.
This quantity is plotted in Fig~\ref{fdt2} and shows that the shorter the distances, the higher is the
effective-temperature, unlike in equilibrium where all length scales are associated with the same temperature.

At this stage, a peculiarity of the zero-frequency effective temperature, namely that it does not vanish
as $K_{neq}\rightarrow K_{eq}$ should be noted. In this case, the first equality in Eq.~(\ref{teff0def})
shows that the effective-temperature is a ratio of two
quantities, both of which go to zero. However, the limit approaches a finite value. The origin of this result that the
temperature approaches a finite value as the quench becomes smaller and smaller is due to the
singular form of the equilibrium distribution function $\coth(\omega/2T)$, which probably persists even for the
weakly nonequilibrium problem. The singular form of $\coth(\omega/2T)$ implies that the limits
$\omega \rightarrow 0$ and $T\rightarrow 0$ do not commute. For one order of limits the answer is divergent,
and for the other it is $1$. In defining the zero-frequency effective-temperature, we have implicitly taken
the frequency to zero first.
This result also
signals that the energy scale $T_{eff,0}$ cannot be used as a good measure of the energy stored in the system due to the
quench. This energy is expected to be distributed in a rather complicated way among all frequency modes.
However, by doing RG we will show that $T_{eff,0}$ acts like
a regular temperature when studying two-point correlation functions as this energy scale causes the
correlations to decay exponentially fast (as compared to a power-law) at long times after the quench.
The size of the quench $|K_0-K|$ on the other hand
is inversely related to how long one has to wait to see this behavior.

Besides an effective-temperature, perhaps a more surprising result, is a generation of friction.
This effect appears at
this stage as a non-zero slope of $2{\rm Im}[C^R_{\phi\phi}] \propto -i\eta \omega$.
In the next section when we do RG,
we will show that $\eta$ corresponds to a finite-lifetime of the bosonic modes, an effect
which is distinct from the generation
of an effective-temperature. Thus we will find that
even though the system we study is closed, and in equilibrium is characterized by long-lived
bosonic modes with $\eta=0$, the quench together with the mode-coupling
arising due to the periodic potential gives rise to additional scattering which generates an $\eta$.
The finite lifetime of
low-frequency bosonic modes implies that there is a flow of energy from low energy scales to high energy scales.
An alternate, but simpler example of this phenomena is the decay of collective modes of a system of
one-dimensional weakly interacting fermions via the creation of particle-hole excitations,
where the fermions are in a nonequilibrium state due to an initial quench.~\cite{Lancaster11}


\section{Derivation of the RG equations} \label{rg}

In order to derive the RG equations, it is convenient to write the Keldysh action for the steady-state,
\begin{eqnarray}
Z_K = \int {\cal D}\left[\phi_{cl},\phi_q\right] e^{i \left(S_0 + S_{sg}\right)}
\end{eqnarray}
where $S_0$ is the quadratic part which describes the physics at long times after the interaction quench,
\begin{eqnarray}
S_0 = \sum_{q,\omega}\begin{pmatrix} \phi_{cl}^*(q,\omega) & \phi_q^*(q,\omega)\end{pmatrix}
&&\begin{pmatrix} 0&&G_A^{-1}\\
G_R^{-1} && -G_R^{-1} G_K G^{-1}_A
\end{pmatrix}\nonumber \\
&&\times\begin{pmatrix}
\phi_{cl}(q,\omega)\\
\phi_q(q,\omega)
\end{pmatrix}\label{S0}
\end{eqnarray}
where,
\begin{eqnarray}
-i\langle \begin{pmatrix}
\phi_{cl}\\\phi_q \end{pmatrix}
\begin{pmatrix} \phi_{cl}^* & \phi_q^*\end{pmatrix} \rangle
= \begin{pmatrix} G_K & G_R \\ G_A & 0\end{pmatrix}
\end{eqnarray}
Using the expressions for $G_{R,A,K}$ in Eqns.~(\ref{GR}) and~(\ref{GK}),
Eq.~(\ref{S0}) may be written in the following manner,
\begin{eqnarray}
&&S_0 = \sum_{q,\omega}\begin{pmatrix} \phi_{cl}^*(q,\omega) & \phi_q^*(q,\omega)\end{pmatrix}
\nonumber \\
&&\frac{1}{\pi Ku}\begin{pmatrix} 0&& (\omega-i\delta)^2 - u^2 q^2       \\
(\omega + i \delta)^2 - u^2 q^2
&& 4i|\omega|\delta\frac{K_0}{2K}\left(1 + \frac{K^2}{K_0^2}\right)
\end{pmatrix} \nonumber \\
&&\times \begin{pmatrix}
\phi_{cl}(q,\omega)\\
\phi_q(q,\omega)
\end{pmatrix}\label{S0a}
\end{eqnarray}
where $\delta=0+$, and the modes are long-lived.
Note that the above action though obtained for the specific case of an interaction
quench in a Luttinger liquid, has the same generic form as any Luttinger liquid subjected to
a nonequilibrium noise source. The details of the noise determines
the particular form of the
coefficient of the $\phi_q^2$-term in the action.~\cite{Torre10,Torrelong} In fact nonequilibrium noise
can also arise in open systems driven out of equilibrium by current flow.~\cite{Mitra06}

We now discuss how the above free theory is affected by a mode-coupling term
due to a periodic potential. The action corresponding to this is
\begin{eqnarray}
&&S_{sg} = \frac{g u}{\alpha^2}\int dx \int dt\left[\cos{\gamma \phi_-}-\cos{\gamma\phi_+}\right]
\end{eqnarray}

We split the fields into slow and fast components,
\begin{eqnarray}
\phi_{cl,q}(xt) = \phi_{cl,q}^{<}(xt)+ \phi_{cl,q}^{>}(xt)
\end{eqnarray}
and integrate out the fast components. We will outline two different procedures for integrating out the
fast modes. In one we impose a hard cutoff, where
the fast fields are defined as,
\begin{eqnarray}
\phi_{cl,q}^{<}(xt) = \int_{-\infty}^{\infty}\frac{d\omega}{2\pi}\int_{-\Lambda^{\prime}}
^{\Lambda^{\prime}}\frac{dq}{2\pi} e^{i q x - i \omega t} \phi_{q,cl}(q,\omega)\label{c1}\\
\phi_{cl,q}^{>}(xt) = \int_{-\infty}^{\infty}\frac{d\omega}{2\pi}\int_{\Lambda > |q| > \Lambda^{\prime}}
\frac{dq}{2\pi} e^{i q x - i \omega t} \phi_{q,cl}(q,\omega)\label{c2}
\end{eqnarray}
and $\Lambda/\Lambda^{\prime} = e^{d\ln(l)}$. The results in Ref.~\onlinecite{Mitra11b}
were presented far from the critical point using the above RG scheme.

The second way of integrating out fast modes is a scheme outlined
by Nozieres and Gallet,~\cite{Nozieres87} where the
slow and fast correlators are defined as follows,
\begin{eqnarray}
G_{0,\Lambda} = G^{<}_{0,\Lambda-d\Lambda} + G^{>}_{\Lambda-d\Lambda,\Lambda}
\end{eqnarray}
Thus the fast correlator may be obtained from taking a derivative of the slow or full correlator,
\begin{eqnarray}
G^{>}_{\Lambda-d\Lambda,\Lambda} = d\Lambda \frac{dG_{\Lambda}}{d\Lambda}
\end{eqnarray}
where $\Lambda = \alpha^{-1}$. Note that when doing RG in real time,
all cutoffs should be imposed only in momentum
space, as imposing cutoffs in time and varying them during the RG flow lead
to inconsistencies such as a violation
of causality.

The Nozieres-Gallet scheme is a more consistent way to deal with the cutoff, and
finally the results in this paper will be presented using this method.
However even the hard cutoff scheme outlined in
Eqns.~(\ref{c1}) and~(\ref{c2}) and used in Ref.~\onlinecite{Mitra11b}
gives qualitatively similar results. Furthermore, in this paper we will use the Nozieres-Gallet
scheme to study the behavior close to the critical point.

Now our task is to expand $Z_K$ in powers of $g$, integrate over the fast modes, and rescale the
cutoff back to its original value ($dx (dt) \rightarrow \frac{\Lambda}{\Lambda^{\prime}} dx (dt)$).
Up to quadratic order in $g$ we find,
\begin{eqnarray}
&&Z_K = \int {\cal D}\left[\phi_{cl}^{<},\phi_q^{<}\right] e^{iS_0^{<}}
\left[1 + i\frac{gu}{\alpha^2} \left(\frac{\Lambda}{\Lambda^{\prime}}\right)^2
\int dx dt\right.\nonumber \\
&&\left.\times e^{-\frac{\gamma^2}{4}\langle (\phi_{cl}^{>})^2\rangle}\left(\cos(\gamma\phi_{-}^{<}(1)) - \cos(\gamma\phi_+^{<}(1))\right)
\right. \label{g1}\\
&&\left. -\frac{g^2u^2}{\alpha^4}\int dx_1 dt_1 dx_2 dt_2\theta(t_1-t_2)\right.\nonumber \\
&&\left.\langle\cos{(\gamma(\phi_{-}^{<}(1)+\phi_-^{>}(1)))} \cos{(\gamma(\phi_{-}^{<}(2)+\phi_{-}^{>}(2)))}
\rangle_{>} \right. \nonumber \\
&&\left. -
\frac{g^2u^2}{\alpha^4}\int dx_1 dt_1 dx_2 dt_2\theta(t_2-t_1)\right.\nonumber \\
&&\left. \langle\cos{(\gamma(\phi_{+}^{<}(1)+\phi_+^{>}(1)))} \cos{(\gamma(\phi_{+}^{<}(2)+\phi_{+}^{>}(2)))}
\rangle_{>} \right. \nonumber \\
&&\left. +
\frac{g^2u^2}{\alpha^4}\int dx_1 dt_1 dx_2 dt_2\left[\theta(t_2-t_1) + \theta(t_1-t_2)\right]\right.\nonumber \\
&&\left. \langle\cos{(\gamma(\phi_{-}^{<}(1)+\phi_-^{>}(1)))} \cos{(\gamma(\phi_{+}^{<}(2)+\phi_{+}^{>}(2)))}
\rangle_{>}
\right]\nonumber \\
\end{eqnarray}
The above may be re-exponentiated using the cumulative expansion
$\langle e^V\rangle\simeq e^{\langle V\rangle
+ \frac{1}{2}\langle V^2 \rangle -\frac{1}{2}\langle V\rangle^2}$ to obtain
\begin{eqnarray}
Z_K = \int {\cal D}\left[\phi_{cl}^{<},\phi_q^{<}\right] e^{iS_0^{<} + i \delta S}
\end{eqnarray}
The expression for $\delta S$ may be simplified by
dropping terms that are proportional to $e^{i\phi_{\pm}(1) + i\phi_{\pm}(2)}$ as they are more
irrelevant than terms such as $e^{i\phi_{\pm}(1) - i\phi_{\pm}(2)}$.
Furthermore, using the fact that
\begin{eqnarray}
\langle \sin{\left[\gamma \phi^>_{a}(1) - \gamma \phi^{>}_b(2)\right]}\rangle = 0
\end{eqnarray}
all the terms containing $\sin$- functions vanish.
Moreover, using $\cos\phi =
:\cos\phi:
e^{-\langle \phi ^2\rangle/2}$ we find,
\begin{eqnarray}
&&\delta S = \frac{gu}{\alpha^2} \left(\frac{\Lambda}{\Lambda^{\prime}}\right)^2
\int dx \int dte^{-\frac{\gamma^2}{4}\langle (\phi_{cl}^{>})^2\rangle}\nonumber \\
&&\times \left(\cos(\gamma\phi_{-}^{<}(1)) - \cos(\gamma\phi_+^{<}(1))\right)
 \label{g2}\\
&&+ i \frac{g^2u^2}{2\alpha^4}\int dx_1 \int dt_1 \int dx_2 \int dt_2\theta(t_1-t_2)\nonumber\\
&&:\cos{(\gamma(\phi_{-}^{<}(1)-\phi_-^{<}(2)))}:e^{-\frac{\gamma^2}{2}
\langle (\phi_{-}(1) -\phi_{-}(2))^2\rangle}\nonumber\\
&&
\left[1
-
e^{-\frac{\gamma^2}{2}\langle (\phi_{cl}^{>})^2\rangle + \frac{\gamma^2}{2}
\langle (\phi_{-}^>(1) -\phi_{-}^>(2))^2\rangle}
\right]  \\
&&
+ i\frac{g^2u^2}{2\alpha^4}\int dx_1 dt_1 dx_2 dt_2\theta(t_2-t_1)\nonumber \\
&&:\cos{(\gamma(\phi_{+}^{<}(1)-\phi_+^{<}(2)))}: e^{-\frac{\gamma^2}{2}
\langle (\phi_{+}(1) -\phi_{+}(2))^2\rangle}
\nonumber \\
&&\left[1
-
e^{-\frac{\gamma^2}{2}\langle (\phi_{cl}^{>})^2\rangle + \frac{\gamma^2}{2}
\langle (\phi_{+}^>(1) -\phi_{+}^>(2))^2\rangle}
\right]
\\
&&-i
\frac{g^2u^2}{2\alpha^4}\int dx_1 dt_1 dx_2 dt_2\left[\theta(t_2-t_1) + \theta(t_1-t_2)\right]
\nonumber \\
&&:\cos{(\gamma(\phi_{-}^{<}(1)-\phi_+^{<}(2)))}:e^{-\frac{\gamma^2}{2}
\langle (\phi_{-}(1) -\phi_{+}(2))^2\rangle}
\nonumber \\
&&\left[1
-
e^{-\frac{\gamma^2}{2}\langle (\phi_{cl}^{>})^2\rangle + \frac{\gamma^2}{2}
\langle (\phi_{-}^>(1) -\phi_{+}^>(2))^2\rangle}
\right]
\end{eqnarray}
where $\langle \left(\phi_{\pm}(1) -\phi_{\pm}(2)\right)^2\rangle$ involves averaging all (both slow
and fast) modes.

Eq.~(\ref{g2}) implies,
\begin{eqnarray}
&&g(\Lambda^{\prime}) = g(\Lambda)\left(\frac{\Lambda}{\Lambda^{\prime}}\right)^2
e^{-\frac{\gamma^2}{4}\langle (\phi_{cl}^{>})^2\rangle}\\
&&= g(\Lambda)\left(\frac{\Lambda}{\Lambda^{\prime}}\right)^2
\exp{\left[-\frac{\gamma^2}{8}K_0(1 + K^2/K_0^2)\ln(\Lambda/\Lambda^{\prime})\right]}\nonumber \\
\label{flowga}
\end{eqnarray}

Note that
\begin{eqnarray}
&&\left[1-
e^{-\frac{\gamma^2}{2}\langle (\phi_{cl}^{>})^2\rangle + \frac{\gamma^2}{2}
\langle (\phi_{a}^>(1) -\phi_{b}^>(2))^2\rangle}
\right]\simeq \nonumber \\
&&
{\cal O}\left(\ln\frac{\Lambda}{\Lambda^{\prime}}\right)
\end{eqnarray}

Next we introduce the center of mass $R = \frac{x_1+x_2}{2}, T_m = \frac{t_1+t_2}{2}$ and
relative coordinates $r = x_1-x_2,\tau=t_1-t_2$, and
expand the ${\cal O}(g^2)$ terms in powers of $r = x_1-x_2, \tau=t_1-t_2$.
Using,
\begin{eqnarray}
&&:\cos{(\gamma(\phi_{-}^{<}(1)-\phi_-^{<}(2)))}:\simeq \nonumber \\
&&1- \frac{\gamma^2}{4}
\left[\left(r\partial_R +\tau\partial_{T_m}\right)\phi_{cl} +\left(r\partial_R +\tau\partial_{T_m}\right)
\phi_q\right]^2\\
&&:\cos{(\gamma(\phi_{+}^{<}(1)-\phi_+^{<}(2)))}:\simeq \nonumber \\
&&1-\frac{\gamma^2}{4}
\left[\left(r\partial_R +\tau\partial_{T_m}\right)\phi_{cl} -\left(r\partial_R +\tau\partial_{T_m}\right)
\phi_q\right]^2\\
&&:\cos{(\gamma(\phi_{-}^{<}(1)-\phi_+^{<}(2)))}:\simeq \nonumber \\
&&1-\frac{\gamma^2}{4}
\left[\left(r\partial_R +\tau\partial_{T_m}\right)\phi_{cl} + 2 \phi_q\right]^2
\end{eqnarray}
we note that the terms with $1$ and purely classical fields cancel. We regroup the remaining terms
and find the following corrections to the quadratic part of the action, (the correction to the
cosine term is already given above in Eq.~(\ref{flowga}))
\begin{eqnarray}
&&S_0 = \int dR \int d(uT_m) \frac{1}{\pi K}\left[\phi_q\left(\partial_R^2 - \partial_{uT_m}^2\right)
\phi_{cl} \right. \nonumber \\
&&\left. + \phi_{cl}\left(\partial_R^2 - \partial_{uT_m}^2\right)
\phi_{q}  + \frac{\delta u}{u}\phi_q\left(\partial_R^2 + \partial_{uT_m}^2\right)
\phi_{cl}\right. \nonumber \\
&&\left. + \frac{\delta u}{u}\phi_{cl}\left(\partial_R^2 + \partial_{uT_m}^2\right)\phi_q
- 2\frac{\eta}{u} \left(\frac{\Lambda}{\Lambda^{\prime}}\right)\phi_q\partial_{uT_m} \phi_{cl}\right.\nonumber \\
&&\left.+ i\frac{4 T_{eff} \eta}{u^2}\frac{K_0}{2K}\left(1 + \frac{K^2}{K_0^2}\right)
\left(\frac{\Lambda}{\Lambda^{\prime}}\right)^2\phi_q^2\right]\label{actioncorr}
\end{eqnarray}

The above implies the following RG equations,
\begin{eqnarray}
&&\frac{d g}{d\ln l} = \left[2 - \frac{\gamma^2}{8}K_0(1 + K^2/K_0^2)\right]g \label{g}\\
&&\frac{d K^{-1}}{d \ln l} = \frac{\pi g^2}{4\alpha^4}\left(\frac{\gamma^2}{2}\right)^2
\frac{K_0}{2}\left(1+\frac{K^2}{K_0^2}\right)I_K\label{K}
\\
&&\frac{1}{Ku}\frac{du}{d\ln l} = \frac{\pi g^2}{4\alpha^4}\left(\frac{\gamma^2}{2}\right)^2
\frac{K_0}{2}\left(1+\frac{K^2}{K_0^2}\right)I_u\label{u}
\\
&&\frac{d\eta}{d\ln l} = \eta + \frac{\pi g^2K u}{2\alpha^4}\left(\frac{\gamma^2}{2}\right)^2
\frac{K_0}{2}\left(1+\frac{K^2}{K_0^2}\right)I_{\eta}\label{diss}\\
&&\frac{d(T_{eff}\eta)}{d\ln l} = 2T_{eff}\eta+ \frac{\pi g^2u^2K^2}{4\alpha^4}\left(\frac{\gamma^2}{2}\right)^2
I_T\label{temp}
\end{eqnarray}
where (defining ${\rm Re}(x) = (x+x^*)/2, {\rm Im}(x) = (x-x^*)/(2i)$)
\begin{eqnarray}
&&I_T =\int_{-\infty}^{\infty} dr\int_{-\infty}^{\infty} dt\nonumber\\
&&{\rm Re}
\left[ e^{-\frac{\gamma^2}{2}\langle (\phi_-(t,r) -\phi_+(0,0))^2\rangle}F^{a/ng}\right]\\
&&I_{\eta} =\int_{-\infty}^{\infty} dr\int_{-\infty}^{\infty} dt \,t \,\nonumber \\
&&{\rm Im}
\left[ e^{-\frac{\gamma^2}{2}\langle (\phi_-(t,r) -\phi_+(0,0))^2\rangle}
F^{a,ng}\right] \\
&&I_K =\int_{-\infty}^{\infty} dr\int_{0}^{\infty} dt \,(r^2-t^2) \nonumber \\
&&{\rm Im}
\left[ e^{-\frac{\gamma^2}{2}\langle (\phi_-(t,r) -\phi_+(0,0))^2\rangle}
F^{a/ng}\right] \\
&&I_u =\int_{-\infty}^{\infty} dr\int_{0}^{\infty} dt \,(r^2+t^2) \nonumber \\
&&{\rm Im}
\left[ e^{-\frac{\gamma^2}{2}\langle (\phi_-(t,r) -\phi_+(0,0))^2\rangle} F^{a/ng}\right]
\end{eqnarray}
where $F^a$ arises due to the hard cutoff scheme and $F^{ng}$ is due to the Nozieres-Gallet scheme.
In particular,
\begin{eqnarray}
&&e^{-\frac{\gamma^2}{2}\langle (\phi_-(r,t)-\phi_+(0,0))^2\rangle} =\nonumber \\
&&e^{-\frac{K_{neq}}{2}\ln\left(\frac{\alpha^2 + \left(t+r\right)^2}{\alpha^2}\right)-
\frac{K_{neq}}{2}\ln\left(\frac{\alpha^2 + \left(t-r\right)^2}{\alpha^2}\right)}\nonumber\\
&&e^{i\left[K_{eq}\tan^{-1}\left(\frac{t+r}{\alpha}\right) + K_{eq} \tan^{-1}
\left(\frac{t-r}{\alpha}\right)
\right]}\\
&&F^a= \frac{1}{2}\left(e^{i\Lambda(t+r)} + e^{i\Lambda(t-r)}\right)\nonumber \\
&&+\frac{i}{2}\left(\frac{K_{eq}}{K_{neq}}-1\right) \left\{\sin(\Lambda(t+r)) + \sin(\Lambda(t-r))\right\}\nonumber \\
\\
&&F^{ng}
=\frac{1}{2}\left[\frac{\alpha^2}{\alpha^2 +(t+ r)^2}
+\frac{\alpha^2}{\alpha^2 + (t - r)^2} \right. \nonumber \\
&&\left. + i\left(\frac{K_{eq}}{K_{neq}}\right)
\left(\frac{\alpha(t+r)}{\alpha^2 + (t+r)^2} + \frac{\alpha(t-r)}{\alpha^2 + (t-r)^2}
\right)\right]\nonumber \\
\end{eqnarray}

\section{Results in the gapless phase}\label{sol}

\subsection{General structure of equations}

In this section we study the consequence of Eqns.~(\ref{g}),~(\ref{K}),~(\ref{u}),~(\ref{diss}) and~(\ref{temp}).
Eq.~(\ref{g})
shows that there is a critical point located at $K_{neq}=\frac{\gamma^2}{8}K_0(1+K^2/K_0^2)=2$,
and therefore at a value of $K$ which is different
from the equilibrium critical point at $K = 8/\gamma^2$. Since $K_{neq}> K_{eq}$, this critical point is always located
at a smaller value of $K$, which is another way of saying that the cosine potential for the post-quench case
is more irrelevant.

Eq.~(\ref{K}) represents the flow of the interaction constant, however depending on
the quench protocol (and hence the values of $K_{eq}$ and $K_{neq}$), the flow can be significantly
different from the
equilibrium flow. For example, while
$I_K >0$ in equilibrium, implying a decrease of the interaction parameter in the presence of the periodic potential,
for the quench problem  $I_K$ can become negative, and also diverge at the critical point. This will be discussed in more
detail below.
Eq.~(\ref{u}) is the flow of the velocity which occurs even in equilibrium,
and is primarily due to the cutoff procedure employed here which does not preserve Lorentz invariance.
The effects of this are small, and in what follows, we will ignore it.

Eq.~(\ref{diss}) shows that a dissipation or a finite lifetime of the bosons $\eta$
is generated, changing the low-frequency properties of the bosonic system qualitatively.
Further, $I_{\eta}$ (which is the rate at which $\eta$ increases with flow)
diverges at the critical point implying a diverging dissipation.
The origin of this divergence is similar to the
divergence of $I_K$ briefly mentioned above. This divergence can have two possible causes:
i) the flow is derived using the
original (athermal state) correlation functions. Since a finite temperature and dissipation is generated,
this might regularize the
divergence close to critical point; ii) the renormalization of the coefficients $\eta$ and $K$
results from a gradient expansion
of the second order results. Such a divergence in the correction might indicate that this expansion breaks down.
This could mean a non-analytical
behavior at low energy. Unraveling this point is a challenging question.

Finally Eq.~(\ref{temp}) shows that a constant term is generated for the strength of the noise correlator
in the zero-frequency limit, which can be interpreted as a product of
dissipation and effective-temperature. In general this noise is expected to have a complicated
frequency dependence
as was highlighted for the $g=0$ case in the previous section. To determine how this
frequency dependence evolves with
RG is a daunting task, and in this paper we will only discuss the effects of $g$ on the
low-frequency part of the noise spectrum.

In both the hard cutoff scheme as well as the Nozieres-Gallet scheme, the important result that
$I_{T,\eta}=0$ when $K_{eq}=K_{neq}$ is recovered. Further both schemes reveal the peculiarity of
diverging $I_K$ and $I_{\eta}$ at
the critical point $K_{neq}=2$ for the nonequilibrium problem.
It should be noted that in deriving the above RG equations we have used the
correlators for $T_{eff}=\eta=0$. The more consistent way to treat the problem is to evaluate the correlators
for non-zero $T_{eff}$ and $\eta$. However this makes the problem quite difficult, all the more so because the
full frequency dependence of $T_{eff}$ is needed.
We argue that taking these effects into account only has a minor influence on the RG flows far
from the critical point.
However the divergences in $I_{\eta,K}$ indicates that
near the critical point, a more consistent computation may be needed. This is clearly something left for future
studies.

In the Nozieres-Gallet scheme, the expressions $I_{T,\eta}$ are found to be
related in a rather simple way to $I_{T,\eta 0}$ {i.e.}, to the $q=0,\omega \rightarrow 0$ limits of the
$C^K$ and ${\rm Im}C^R$ evaluated for the quadratic theory in Section~\ref{quench}. In particular,
\begin{eqnarray}
I_T &&= \frac{8\alpha^2}{K_{neq}(K_{neq}-1)}\nonumber \\
&&\times \left[\frac{\pi}{2^{K_{neq}}
B\left(\frac{K_{neq}+K_{eq}}{2},
\frac{K_{neq}-K_{eq}}{2}\right)}\right]^2\nonumber \\
&&= \left(\frac{K_{neq}-1}{K_{neq}}\right)I_{T0}
\end{eqnarray}
\begin{eqnarray}
I_{\eta} &&= 8\alpha^3
\left[\frac{\pi}{2^{K_{neq}}
B\left(\frac{K_{neq}+K_{eq}}{2},
\frac{K_{neq}-K_{eq}}{2}\right)}\right]^2\nonumber \\
&&\times \left(\frac{K_{eq}}{K_{neq}}\right)\frac{(K_{neq}-3/2)}{(K_{neq}-1)^2(K_{neq}-2)}\nonumber \\
&&= \left(\frac{K_{neq}-3/2}{K_{neq}}\right)I_{\eta 0}
\end{eqnarray}
where $I_{T\eta 0}$ are given in Eqns.~(\ref{It0}) and~(\ref{Ieta0}).
The ratio of these quantities leads to an effective-temperature,
\begin{eqnarray}
&&T_{eff}^{ng}=\frac{\alpha I_T}{2I_{\eta}} = \frac{(K_{neq}-1)(K_{neq}-2)}{2K_{eq}(K_{neq}-3/2)}\\
&&=\left(\frac{K_{neq}-1}{K_{neq}-3/2}\right){T}_{eff,0}\label{tint}
\end{eqnarray}
with $T_{eff,0}$ being the non-interacting expression for the effective-temperature. Note that when $K_{neq}\gg 1$,
the cosine potential is more irrelevant, and the above expressions approach the non-interacting values. The above
analytic expressions also show that
$I_{\eta}$ is divergent at the critical point as $1/(K_{neq}-2)$.

\begin{figure}
\includegraphics[totalheight=5cm
]{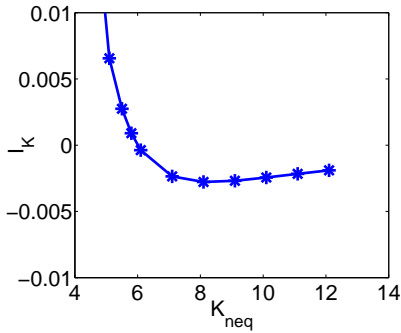}
\caption{\label{IKsign} Change of sign of $I_K$ at a certain value of $K_{neq}$. Here $K_{eq}=2$}
\end{figure}

The expression for $I_K$ is more complex and given by,
\begin{eqnarray}
&&I_K =-\alpha^4\left[\int_0^{\pi/2}dx\tan{x}\cos(K_{eq}x)(\cos{x})^{K_{neq}}
\right. \nonumber \\
&&\left. \times \int_0^x dy \tan{y} \sin(K_{eq}y)(\cos{y})^{K_{neq}-2}\right. \label{ik1} \\
&&\left. + \int_0^{\pi/2}dx\tan{x}\cos(K_{eq}x)(\cos{x})^{K_{neq}-2}\right.\nonumber \\
&&\left. \times \int_0^x dy \tan{y} \sin(K_{eq}y)(\cos{y})^{K_{neq}}\right. \label{ik2} \\
&&\left. + \left(\frac{K_{eq}}{K_{neq}}\right)
\int_0^{\pi/2}dx\tan{x}\cos(K_{eq}x)(\cos{x})^{K_{neq}-2}\right. \nonumber \\
&&\left. \times \int_0^x dy (\tan{y})^2 \cos(K_{eq}y)(\cos{y})^{K_{neq}}\right. \label{ik3}\\
&&\left. -\left(\frac{K_{eq}}{K_{neq}}\right)\int_0^{\pi/2}dx(\tan{x})^2\sin(K_{eq}x)
(\cos{x})^{K_{neq}}\right.\nonumber \\
&&\left. \times \int_0^x dy \tan{y} \sin(K_{eq}y)(\cos{y})^{K_{neq}-2} \right]\label{ik4}
\end{eqnarray}
$I_K$ shows complicated structure, and in particular depending on the quench protocol can be positive,
negative or zero.
For example, $I_K(K_{eq}=2,K_{neq}=7)=-\alpha^4 0.0022$ showing that for this quench protocol,
the periodic potential
increases the interaction parameter, making the system more delocalized as compared to the
ground state of a Luttinger liquid with interaction parameter $K$.
A plot of $I_K$ for $K_{eq}=2$ and different $K_{neq}$ is shown in
Fig~\ref{IKsign} where a change of sign of $I_K$ is
found.

Further, $I_K$ generically diverges at the critical point $K_{neq}=2$ for all values of $K_{eq}$
except $K_{eq}=1$ and $K_{eq}=2$.
This divergence is seen from observing that when $K_{neq}=2$, the Eqns.~(\ref{ik2})
and~(\ref{ik3}) above are in general divergent because $\tan(\pi/2)=\infty$ and
$(\cos(x))^{K_{neq}-2}=1$. However these divergences cancel in equilibrium when $K_{eq}=K_{neq}=2$, and we
find $I_{K}(K_{eq}=K_{neq}=2)= \alpha^4(8\pi/32)$. In contrast at the
special point $K_{eq}=1,K_{neq}=2$, $I_K(K_{eq}=1,K_{neq}=2)=0$.

In the subsequent sections we will study the RG flows in the gapless phase for two cases, one where the system is
far from the critical point, and the other where it is close to it. For the latter we will highlight the
effect of the above divergences.

\subsection{Flow far from the critical point}

\begin{figure}
\includegraphics[totalheight=5cm]{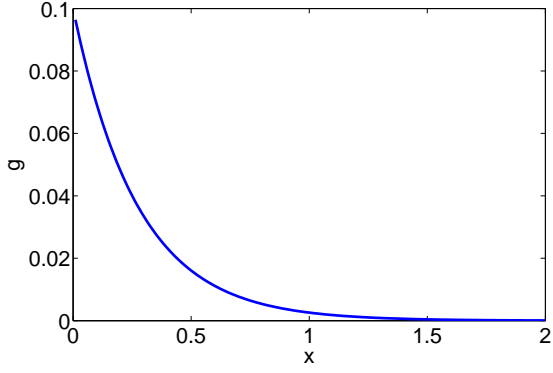}
\caption{\label{gflow} Flow of $g$ with $x = \ln{l}$ for $K_0=3$, $K=5$, $g=0.1$ and $\gamma=2$.}
\end{figure}

\begin{figure}
\includegraphics[totalheight=5cm
]{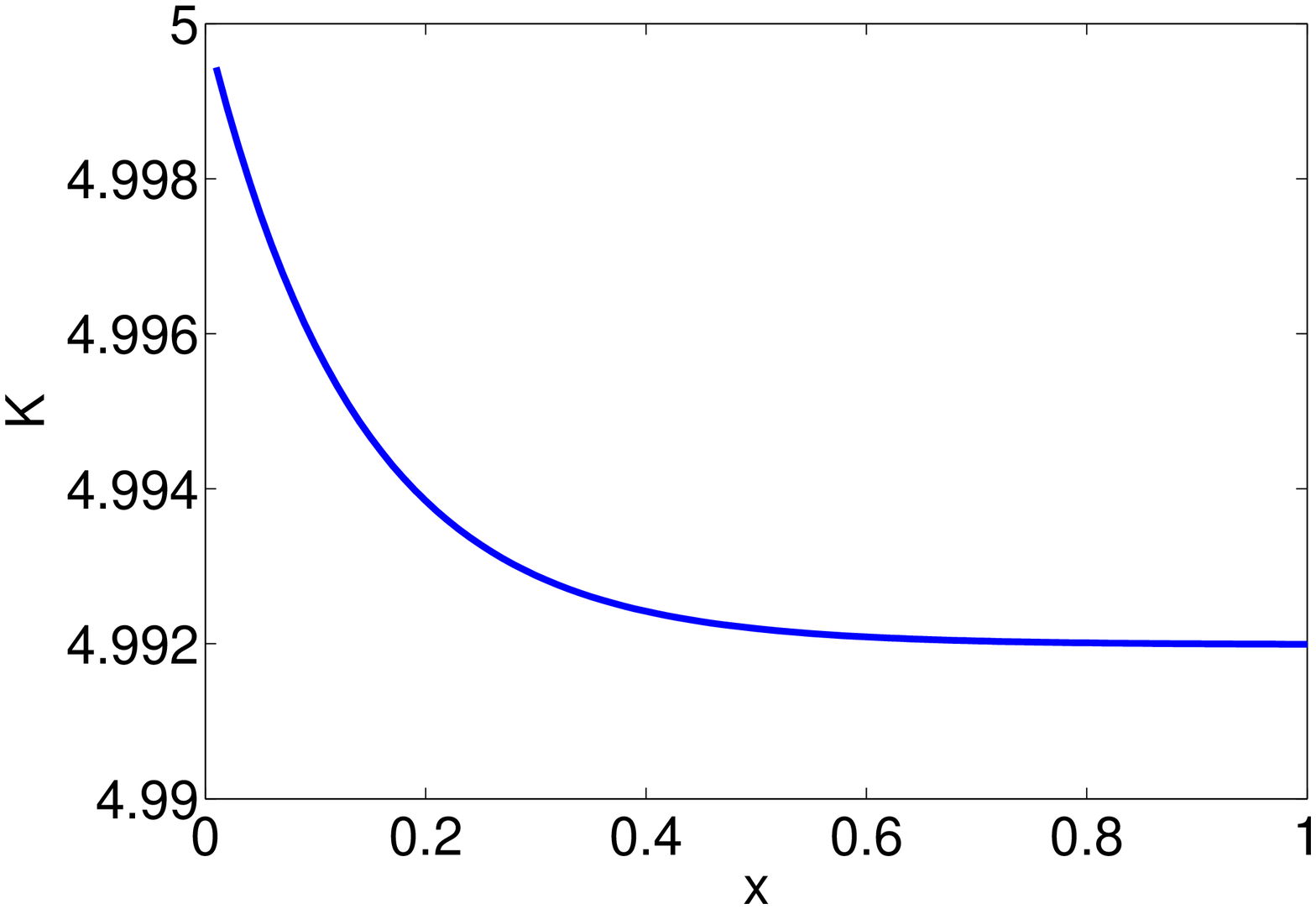}
\caption{\label{kflow} Flow of $K$ with $x = \ln{l}$ for $K_0=3$, $K=5$, $g=0.1$ and $\gamma=2$.}
\end{figure}

\begin{figure}
\includegraphics[totalheight=5cm
]{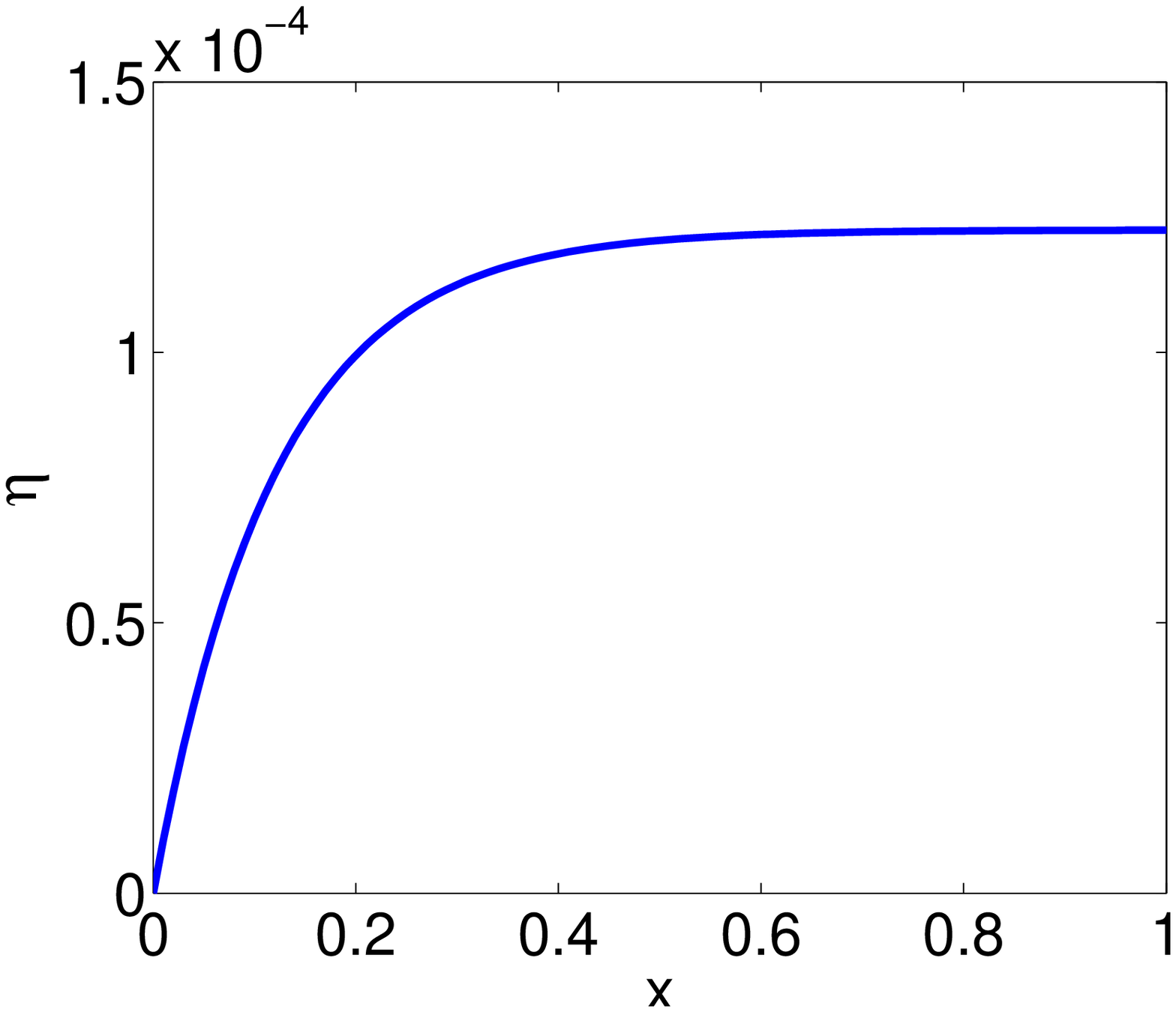}
\caption{\label{dissflow} Flow of $e^{-x}\eta(x)$ with $x = \ln{l}$ for $K_0=3$, $K=5$, $g=0.1$ and $\gamma=2$.}
\end{figure}

\begin{figure}
\includegraphics[totalheight=5cm
]{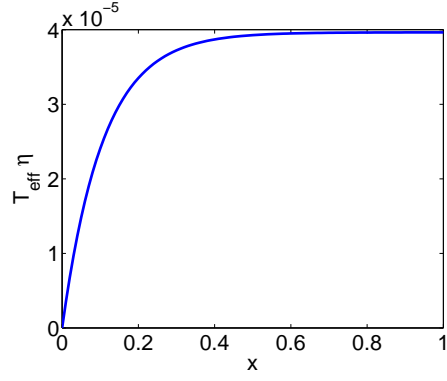}
\caption{\label{tempdissflow}Flow of $e^{-2x}T_{eff}\eta $ with $x = \ln{l}$ for $K_0=3$, $K=5$, $g=0.1$
and $\gamma=2$.}
\end{figure}

The RG equations can be solved numerically. The flow of the various quantities
for a quench from $K_0=3$ to $K=5$ and for $\gamma =2$ are shown in
Figs.~\ref{gflow},~\ref{kflow},~\ref{dissflow} and~\ref{tempdissflow}. For these values
$I_K$ remains positive, so that $K$ decreases during the flow (Fig.~\ref{kflow}).
Since $g$ is (dangerously) irrelevant it flows to zero (Fig.~\ref{gflow}), however for the nonequilibrium problem,
$g$ generates new terms such as a dissipation (Fig.~\ref{dissflow}) and a zero-frequency component of the
noise (Fig.~\ref{tempdissflow}).
The latter may be identified from the zero frequency limit of the following term in equilibrium,
\begin{equation} \label{eq:deftemp}
\eta \omega \coth(\omega/2T) \to 2 \eta T
\end{equation}
as ${\rm dissipation}\times {\rm effective-temperature}$.
Note that since we have, from the RG, a quadratic action for the fixed point, this allows us to unambiguously
define a temperature for the \emph{low energy} modes. Indeed with such a quadratic action the
fluctuation-dissipation
theorem would be obeyed for all correlation functions involving the field $\phi$ with the same temperature $T$
as defined by (\ref{eq:deftemp}). One thus sees that the RG procedure indicates that the system does thermalize.
It is important to remember however that this statement only concerns the low energy modes. The frequency
dependence of the Keldysh term is quite different from the standard $\eta \omega \coth(\omega/2T)$.
This means that as the frequency is increased there
will be a complicated crossover between this thermal state at low energy and the athermal original state
that controls the high energy behavior of the system. Another interesting question is what will happen
for the correlation functions of the dual field $\theta$. The most naive expectation would be that
such correlations
are also controlled by the same temperature $T$.
However since in the above action $\theta$ has been integrated out,
this would need explicit calculations of the $\theta-\theta$ correlations, a quite complicated calculation.
It will thus be interesting to see whether this is indeed the case or whether the dual field
can have a different behavior.

The flow of the effective-temperature
can be studied most easily by substituting Eq.~(\ref{diss}) in Eq.~(\ref{temp}). We obtain,
(after defining dimensionless variables $T_{eff} \rightarrow \alpha T_{eff}/u, \eta \rightarrow \alpha \eta/u$)
\begin{eqnarray}
&&\frac{d T_{eff}}{d\ln{l}} = T_{eff} + \frac{\pi g^2}{4}\left(\frac{\gamma^2}{2}\right)^2
\frac{K^2 I_T}{\eta}\nonumber \\
&&\times \left[1 - 2 T_{eff}\left(\frac{K_0}{2K}\right)\left(1+\frac{K^2}{K_0^2}\right)
\frac{I_{\eta}}{I_T}
\right]
\end{eqnarray}
Note that the physical temperature $T_{eff}e^{-\ln{l}}$ obeys the differential equation
\begin{eqnarray}
&&\frac{d \left(T_{eff}e^{-\ln{l}}\right)}{d\ln{l}} =
e^{-\ln{l}}\frac{\pi g^2}{4}\left(\frac{\gamma^2}{2}\right)^2
\frac{K^2 I_T}{\eta}\nonumber \\
&&\left[1 - 2 T_{eff}\left(\frac{K_0}{2K}\right)\left(1+\frac{K^2}{K_0^2}\right)
\frac{I_{\eta}}{I_T}
\right]
\end{eqnarray}
The above shows that the steady-state solution corresponds to a temperature,
\begin{eqnarray}
T_{eff}^* =\left(\frac{K_{neq}^*-1}{K_{neq}^*-3/2} \right)\left(\frac{K_{neq}^*-2}{2K_{neq}^*}\right)
\label{tsteady}
\end{eqnarray}
Note that $T_{eff}^*$ differs from the value of $I_T/(2I_{\eta})$ in Eq.~(\ref{tint}) only because of
the way we have
defined it. The energy-scale which determines the decay of the correlations is the prefactor of the
$\phi_q^2$ term in the action Eq.~(\ref{actioncorr}), and is the combination $T_{eff}^* K_{neq}^*/K_{eq}^*$.


\begin{figure}
\includegraphics[totalheight=5cm
]{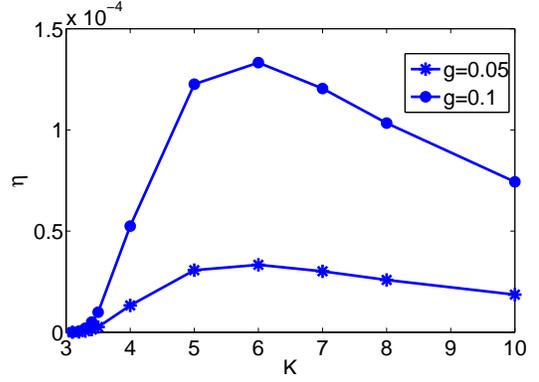}
\caption{\label{dissfig1} Strength of the dissipation $\eta$ for $K_0$=$3$, $\gamma$=$2$ and $g$=$0.05$ and
$g$=$0.1$.}
\end{figure}

The renormalized dissipation $e^{-\ln l}\eta$ for a quench from $K_0=3$ to $K > K_0$ is shown in
Fig.~\ref{dissfig1}.
The behavior is non-monotonic as the size of the quench becomes larger and larger.
The reason for this is that when $K=K_0$, $\eta=0$. At the same time when $K \gg K_0$, the
cosine potential being more irrelevant, decays faster to zero, so that the renormalized $\eta$ is also smaller
for larger $K$. These two behaviors for $|K-K_0|\ll 1$ and $K \gg K_0$ imply a maxima in between.

\begin{figure}
\includegraphics[totalheight=7cm]{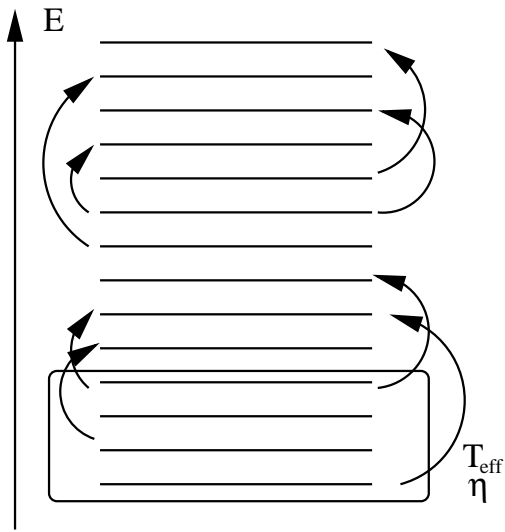}
\caption{\label{cartoon} A schematic of the RG procedure. The resultant dissipation and temperature generated
in the low-energy theory (box) arises due to an exchange of energy with the high-energy degrees of freedom
that act as a bath.}
\end{figure}

The generation of a temperature and dissipation in the low-energy theory via the RG procedure is shown
schematically in Fig.~\ref{cartoon}. For the out of equilibrium system, there is a gradual
flow of energy from the low-energy (long wavelength) degrees of freedom and the high-energy (short wavelength)
degrees of freedom. Thus when the latter are integrated out in the RG procedure,
this flow of energy appears as a dissipation in the low-energy sector. Such a dissipation is also
accompanied by a temperature such that a low-frequency classical FDT is obeyed.
The above mechanism for thermalization arising due to an exchange of energy between long
wavelength and short wavelength modes appears to be rather generic, and may even be recovered in
a quench involving fermions.~\cite{Lancaster11}
In particular in Ref.~\onlinecite{Lancaster11}
the effect of weak interactions on a system of one-dimensional fermions that are in a
nonequilibrium state due to an initial quench was studied using the random phase approximation (RPA).
The RPA analysis revealed that the highly
nonequilibrium fermion distribution generated by the quench results in an
enhanced particle-hole continuum. Thus for attractive interactions between fermions
the collective modes were found to lie within this continuum, and were therefore
found to be overdamped.

It would be interesting to study other models with non-linearities to see if a similar generic
fixed point is reached, and also how the temperature might depend on the type of non-linearities
that exist in the Hamiltonian. Among the various non-linearities that
one could think of, there exists in particular the band curvature that is inherent to a
realistic one dimensional system. Such a band curvature
leads in particular to $(\nabla\phi)^3$ terms. Whether such terms could lead to similar effects is
an interesting question.
Note that in equilibrium, because of conservation laws,
the cosine terms are more efficient in mixing certain modes in contrast to other non-linearities.
In particular the curvature terms were found not to be able to
fully relax a current in Ref~\onlinecite{giamarchi_curvature}.
Whether similar differences also exist out of equilibrium is clearly a challenging question.

Independently of these subtle points we think that the mechanism that we could obtain
in a controlled way for this particular
model using the RG procedure is quite generic. Fig.~\ref{cartoon}
indicates that provided enough mode coupling exists in a system,
the system itself can act as a reservoir and bath for the low energy part of the degrees of freedom.
This sub-part will thus acquire a classical behavior in the sense that it will get a finite dissipation
and a finite temperature. It is important to contrast this with a standard
equilibrium quantum system for which the temperature is the same irrespectively of the energy of the mode
considered. Here the temperature can
only be defined if the limit $\omega \to 0$ is taken. The frequency dependence will in general be
complicated and correspond to the crossover between the athermal and the (low energy) thermal state.

\subsection{Flow in the vicinity of the critical point}

\begin{figure}
\includegraphics[totalheight=5cm
]{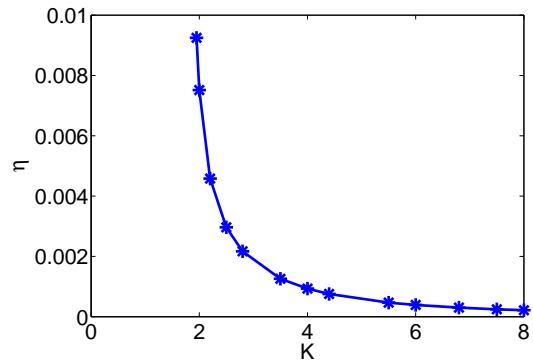}
\caption{\label{dissfig2} Strength of the dissipation $\eta$ for $g=0.05$, $K_0=1$ and $\gamma=2$ so
that $K_{neq} = \frac{1}{2}(1+K^2)$. The
critical point is located at $K = \sqrt{3}$.}
\end{figure}

In this section we study the RG equations near the critical point and in particular highlight the effects of
diverging $I_{K,\eta}$.
Fig.~\ref{dissfig2} shows the renormalized dissipation for $\gamma=2$ and for a quench
from $K_0=1$ and a $K$ chosen to be $K > \sqrt{3}$. Thus these quenches include those that go across the
equilibrium critical point at $K=2$, and are always on the gapless side of the new nonequilibrium critical point
located at $K =\sqrt{3}$.
Fig.~\ref{dissfig2} shows that as $K$ is varied such that one approaches this new critical point,
the dissipation diverges due to diverging $I_{\eta}$.
This could either signal a breakdown in the gradient expansion
for the nonequilibrium problem and/or a drawback of our approximation of setting $T_{eff}=\eta=0$
in the two-point functions
used in the evaluation of the RG equations.

Neglecting the effects of temperature and dissipation, it is interesting to study the
BKT flow near the quantum critical point. It is of course understood that the flow would be eventually
cutoff by the temperature and/or the dissipation.
Note that close to the quantum critical point $K_{neq} = 2$, the temperature (\ref{tsteady}) becomes
parametrically small while $\eta$ diverges, so that the product $\eta T_{eff}$ is well defined.
Yet, the parametrically small temperature, probably signals
that as one approaches the Mott-phase, the low energy theory is no longer described by gapless thermalized modes.
This is an important open question which we do not address in this paper.

\begin{figure}
\includegraphics[totalheight=5cm,]{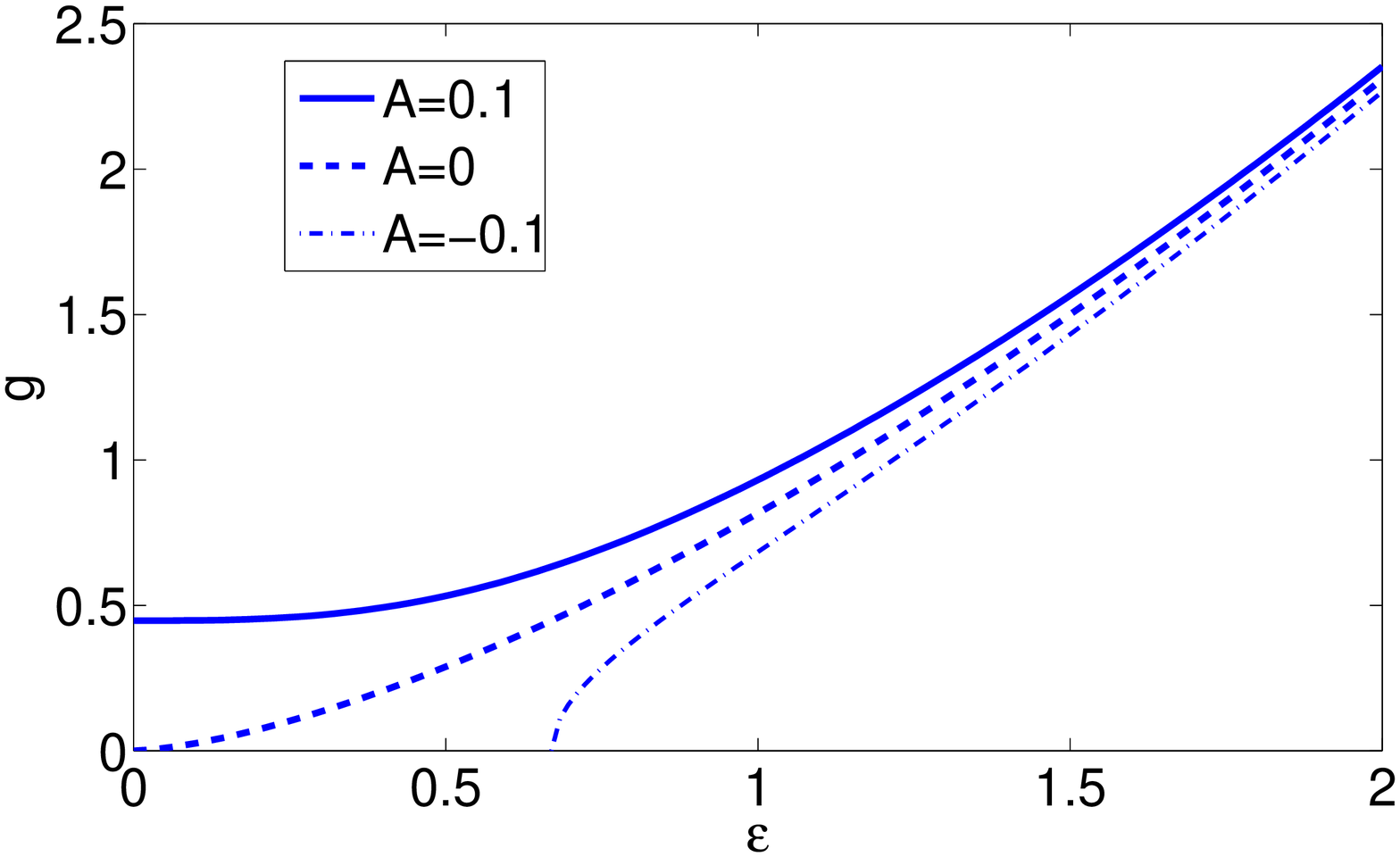}
\caption{\label{ktneq} Flow near the critical point $K_{neq}=2 +\epsilon$, neglecting the effects of temperature and
dissipation. Note that the boundary separating the regions in which the cosine is relevant or not is quite different
from the equilibrium BKT one.}
\end{figure}

Near the critical point we may
expand the RG equations for $g$ and $K$ (Eqns.~(\ref{g}) and~(\ref{K})) about $K_{neq} = 2 +\epsilon$. For concreteness
let us set $K_{eq}=3/2$. Then the complicated expression for $I_K$ in
Eqns.~(\ref{ik1}),~(\ref{ik2}),~(\ref{ik3}) and~(\ref{ik4})
reduces to
\begin{eqnarray}
&&I_K(K_{eq}=3/2,K_{neq}=2+\epsilon) = -\alpha^4\left[-0.54 \right. \nonumber \\
&&\left. + \frac{4}{7}\int_0^{\pi/2}dx \tan{x}\left(\cos{x}\right)^{\epsilon}\left(\sin\frac{x}{2}\right)^3\right.\nonumber\\
&&\left. \times \left(\cos{\frac{3x}{2}}\right)\left(3 + 3 \cos{x} + \cos{2x}\right)
\right. \nonumber \\
&&\left. + \frac{1}{7}\int_0^{\pi/2}dx \tan{x}\left(\cos{x}\right)^{\epsilon}\left(\sin\frac{x}{2}\right)^3\right.\nonumber\\
&&\left. \times \left(\cos{\frac{3x}{2}}\right)\left(2 + 9 \cos{x} + 3 \cos{2x}\right)
\right]
\end{eqnarray}
where the first smooth numerical term comes from Eqns,~(\ref{ik1}) and~(\ref{ik4}), while the rest are terms that diverge as $\epsilon\rightarrow 0$.
Thus for $\epsilon\ll 1$, we may approximate
\begin{eqnarray}
I_K \simeq \frac{c}{\epsilon}
\end{eqnarray}
where $c$ is a positive number. Thus after a redefinition
of $g$, the flow of $g$ and $\epsilon$ are found to be
\begin{eqnarray}
\frac{dg}{d\ln{l}} = -\epsilon g\\
\frac{d\epsilon}{d\ln{l}} = -\frac{g^2}{\epsilon}
\end{eqnarray}
The above imply that the flow equations are along the following lines
\begin{eqnarray}
\frac{g^2}{2}-\frac{\epsilon^3}{3} = A
\end{eqnarray}
where $A$ is a constant. Some examples for different choices of $A$ are shown in Fig.~\ref{ktneq} and should be
contrasted with the equilibrium BKT flow which are along the lines $g^2 - \epsilon^2 = {\rm constant}$.

Along the separatrix $A=0$, the solution of the RG equations give
\begin{eqnarray}
\epsilon(l) = \frac{\epsilon_0}{1+\frac{2}{3}\epsilon_0\ln{l}}\\
g(l) = g_0\left[\frac{1}{1+\left(\frac{2g_0}{3}\right)^{2/3}\ln{l}}\right]^{3/2}
\end{eqnarray}
(denoting $\epsilon_0,g_0$ as the initial bare values).

\begin{figure}
\includegraphics[totalheight=5cm]{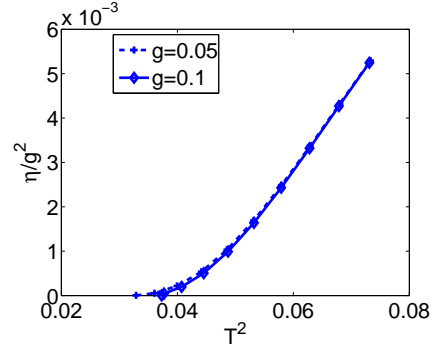}
\caption{\label{fermi} A plot of $\eta/g^2$ vs $T^2$ where $T=T_{eff}^*K_{neq}^*/K_{eq}^*$ and
for quenches where $\gamma=2$,
$K_0=3$ and $K > 3$. }
\end{figure}

\section{Properties of the fixed point action} \label{action}

In this section we discuss the general properties of
the steady-state resulting from the RG. The renormalized action is a quadratic theory of
thermal bosons with a finite lifetime. In the low-frequency limit, the effective action is given by
\begin{eqnarray}
S_0 &&= \sum_{q,\omega}\begin{pmatrix} \phi_{cl}^*(q,\omega) & \phi_q^*(q,\omega)\end{pmatrix}
\nonumber \\
&&\frac{1}{\pi Ku}\begin{pmatrix} 0&& \omega^2-i\eta\omega - u^2 q^2       \\
\omega^2 + i \eta\omega - u^2 q^2
&& 4iT_{eff}\eta\frac{K_0}{2K}\left(1 + \frac{K^2}{K_0^2}\right)
\end{pmatrix}\nonumber \\
&&\begin{pmatrix}
\phi_{cl}(q,\omega)\\
\phi_q(q,\omega)
\end{pmatrix}\label{S0ren}
\end{eqnarray}
Thus the $\langle \phi \phi\rangle$ correlators are,
\begin{eqnarray}
&&G^R(q,\omega) = \frac{\pi K u}{\omega^2-u^2q^2 + i\eta \omega}\\
&&G^K(q,\omega) = (-2i\pi)\left(\frac{K_0}{2K}\right)\left(1+\frac{K^2}{K_0^2}\right)\nonumber\\
&&\times \left(2T_{eff}\eta\right)\frac{uK}{\eta^2\omega^2 +(\omega^2-u^2q^2)^2}
\end{eqnarray}
The above correlators should be contrasted with those of the nonequilibrium
Luttinger liquid discussed in section III. While both effective theories are quadratic,
the combined effect of a quench and the
periodic potential is to give rise to inelastic scattering that
broadens the bosonic modes by an amount given by $\eta$. For the Luttinger
liquid on the other hand (section III), the bosonic modes are long lived,
but are characterized by a nonequilibrium occupation probability.
Note that a finite lifetime $\eta$
may also be generated for interacting bosons in
a periodic potential which is in equilibrium but at a non-zero temperature,
provided there are no special conservation
laws.~\cite{Giamarchi91,Sirker11}

In what follows
we neglect the $\omega^2$ term relative to the $\eta\omega$ term as we are
primarily interested in the long-distance and long-time limits.
Then the retarded correlator is found to be,
\begin{eqnarray}
&&-i\langle \phi_{cl}(1)\phi_q(2)\rangle \nonumber \\
&&= \int \frac{dq}{2\pi}\int \frac{d\omega}{2\pi}
e^{iq(x_1-x_2)-i\omega(t_1-t_2)}G^R(q,\omega)\nonumber \\
&&= -\theta(t_1-t_2)\frac{\sqrt{\pi}}{2}\frac{K}{\sqrt{\eta (t_1-t_2)}}
e^{-\frac{\eta(x_1-x_2)^2}{4u^2|t_1-t_2|}}
\end{eqnarray}
Similarly the advanced correlator is,
\begin{eqnarray}
&&-i\langle \phi_{q}(1)\phi_{cl}(2)\rangle \nonumber \\
&&= \int \frac{dq}{2\pi}\int \frac{d\omega}{2\pi}
e^{iq(x_1-x_2)-i\omega(t_1-t_2)}G^A(q,\omega)\nonumber \\
&&= -\theta(t_2-t_1)\frac{\sqrt{\pi}}{2}\frac{K}{\sqrt{\eta (t_2-t_1)}}e^{-\frac{\eta(x_1-x_2)^2}{4u^2|t_1-t_2|}}
\end{eqnarray}
while the Keldysh correlator is found to be
\begin{eqnarray}
&&-i\left[\langle \phi_{cl}(1) \phi_{cl}(2)\rangle -\langle \phi_{cl}^2\rangle\right]\nonumber \\
&&=
 \int \frac{dq}{2\pi}\int \frac{d\omega}{2\pi}
\left[e^{iq(x_1-x_2)-i\omega(t_1-t_2)}-1\right]G^K(q,\omega)\nonumber \\
&&=i\left(\frac{K_0}{2K}\right)\left(1+\frac{K^2}{K_0^2}\right)\left(2T_{eff}K\right)
\nonumber \\
&&\times \left[\sqrt{\pi \frac{|t_1-t_2|}{\eta}}e^{-\frac{\eta(x_1-x_2)^2}{4u^2|t_1-t_2|}}\right.\nonumber \\
&&\left. -\frac{\pi}{2}\left(\frac{x_1-x_2}{u}\right)Erf\left(\frac{\sqrt{\eta}|(x_1-x_2)|}
{2u\sqrt{|t_1-t_2|}}\right)\right]
\end{eqnarray}
For $t_1=t_2$, the above reduces to
\begin{eqnarray}
&&-i\left[\langle \phi_{cl}(x_1,t) \phi_{cl}(x_2,t)\rangle -\langle \phi_{cl}^2\rangle\right]\nonumber \\
&&=i\left(\frac{K_0}{2K}\right)\left(1+\frac{K^2}{K_0^2}\right)\left(2T_{eff}K\right)
\left[
-\frac{\pi}{2}\left(\frac{x_1-x_2}{u}\right)\right]\nonumber \\\label{phfp}
\end{eqnarray}
Eq.~(\ref{phfp}) shows that the equal-time two-point function $C^K_{\phi\phi}$ decays exponentially in position, with a decay rate given by
the effective-temperature, $T_{eff}K_{neq}/K_{eq}$
\begin{eqnarray}
\langle e^{i\phi_{cl}(x)}e^{-i\phi_{cl}(y)}\rangle\simeq e^{-\frac{T_{eff} K_{neq}}{K_{eq}}\frac{K}{u}|x-y|}
\end{eqnarray}

An interesting question concerns the relation between the generated dissipation and the effective-temperature.
For a fermi-liquid for example  $\eta \sim g^2 T$. It is interesting to explore to what extent our system
mimics such a behavior. Fig.~\ref{fermi} shows the plot of $\eta/g^2$ as a function of $T^2$
where $T = T_{eff}^* K_{neq}^*/K_{eq}^*$,
the latter being the appropriate energy scale that determines the decay of the two-point functions.
The $*$ indicates
that the values at the fixed point have been taken.
Fig.~\ref{fermi} corresponds to $\gamma=2$ and a quench
from $K_0=3$ to $K \geq K_0$. Notice the coincidence of the plots for two different values of $g$
indicating that the dissipation scales as $g^2$. However, for small quenches, the behavior is
not fermi-liquid like as the dissipation increases as $T^{\beta}$ with $\beta > 2$.
For larger quenches, the relation between $\eta$ and $T$ becomes non-monotonic (not shown) as $T$
always increases with $K$,whereas $\eta$ eventually decreases as $g$  becomes more and more irrelevant.

\section{Conclusions and Outlook}\label{summ}

In this paper we have explored how out of equilibrium quantum systems can thermalize in the presence of
mode-coupling terms.
We have analyzed in detail a situation in which a system of interacting bosons is set out of equilibrium by a sudden
interaction quench. In Ref.~\onlinecite{Mitra11b} and the current paper we have shown, using a controlled
RG procedure,
how mode-coupling or interactions which give rise to nontrivial scattering between modes affect this
nonequilibrium state.

The main result is that even when the mode-coupling is ``irrelevant'', it generates an
effective-temperature and a dissipation. The generation of a dissipation indicates that thermalization
eventually sets in by the exchange of energy
between low energy modes and high energy modes. The flow of energy across different length scales
is also found in nonequilibrium classical systems such
as turbulence~\cite{Frisch} and the classic Fermi-Pasta-Ulam problem.~\cite{fpu}
Such a flow of energy can lead to cascades and universal
power-law behavior in the distribution function, a result which has also been recovered in
recent studies involving nonequilibrium Bose-Einstein condensates.~\cite{Gasenzer11}
It is clear that exploring further the connection between these systems should
prove to be a very fruitful line of study.
This also points to quite a general mechanism for thermalization in which mode coupling allows the system itself to
act as a reservoir for the low energy part of the degrees of freedom (see Fig.~\ref{cartoon}).
This subpart thus acquires a classical behavior in the sense that it is characterized by a finite dissipation and
a finite temperature. It is important to contrast this to a standard
equilibrium quantum system for which the temperature is the same irrespective of the energy of the mode considered,
here the temperature can only be defined if the limit $\omega \to 0$ is taken.
The frequency dependence will in general be complicated and will correspond to a crossover between
the athermal and the (low energy) thermal state.

It is also interesting to compare the approach we have employed with
more traditional methods to study dynamics. When studying fermions, a natural route is to
write a kinetic equation that systematically takes into
account two and if needed three particle scattering processes.
However for the Luttinger liquid which has
a linearized spectrum, a description in terms of a kinetic theory fails.
In particular a naive perturbation theory about the Luttinger liquid
fixed point leads to divergent results.
This has been attempted in the past (see
Ref.~\onlinecite{Imambekovrev11} for a discussion on this point).
While our approach is perturbative in the periodic potential, since
the basic unit of our perturbation theory involves the
correlators $\langle e^{i a \phi} e^{-i a \phi}\rangle $, we have explictly taken into account
multiple scattering between bosons.

There are of course many open problems. It is in particular important to extend the analysis of the present paper
close to the critical point and also to the region in which the mode coupling term is perturbatively relevant.
Although we could obtain in the present paper some results close to the critical point, the whole RG procedure
has to be made fully consistent by taking into account the temperature and dissipation to obtain a more
complete description. Treating these effects in a self-consistent way is hard. Once such a scheme is developed,
it would be interesting to
study the strong-coupling part of the phase diagram. It is clear that numerical studies of these
questions would also
be extremely helpful in this regime.

Given the close analogies between fermions and bosons in 1D, exploring the above physics in fermionic systems is
an important direction. A first step was undertaken in Ref.~\onlinecite{Lancaster11} where an RPA study of a
system of fermions
that are out of equilibrium due to an initial quench was done, and overdamped collective modes were recovered.
However,
a more complete study is needed that takes into account backscattering interactions
(which are not included in the RPA).
Further, we have assumed that the term that generates mode-coupling was switched on very slowly. How the physics
is affected by the rate at which the mode-coupling is turned on is also interesting to explore.

Finally checking the generality of this mechanism, by introducing other non-linear terms,
such as those arising due to band-curvature, is of course a very interesting
question. This is also directly relevant for a test of the present mechanism either in
numerical simulations or in experiments. On a practical point there is also the question
of time-scales and which non-linear couplings are more efficient than others.
In experiments in David Weiss's group,~\cite{Kinoshita06} where there is currently no optical
potential, a prethermalized GGE type state is found to persist
for long times despite non-linearities arising due to effects such as band-curvature.
This might be because the time-scales for the onset of dissipative and
thermal effects due to these couplings are too long for experimental relevance. However we expect
that application of a periodic potential will induce these effects more
efficiently as a periodic potential is (dangerously) irrelevant. Once an experiment with an optical
lattice is realized, a probe of the density-density response functions which directly correspond to the
correlators $\langle e^{i a \phi} e^{-i a \phi}\rangle$ that we have evaluated, should
exhibit dissipative and thermal effects.

{\sl Acknowledgements}:
The authors thank  Ehud Altman and Emanuele Dalla Torre for helpful discussions,
and are particularly grateful to Boris Altshuler for pointing out the
connection with turbulence. AM also thanks Institut Lau Langevin and
Aspen Center for Physics for hospitality where part of this work was completed.
This work was supported by NSF-DMR (Grant No. 1004589) (AM)
and by the Swiss SNF under MaNEP and Division II (TG).


%

\end{document}